\documentclass[prd,aps,twocolumn,nofootinbib,amssymb,floatfix,preprintnumbers,superscriptaddress, noeprint]{revtex4-1}
\usepackage{graphicx,amsmath,mathtools,physics}
\usepackage{epsfig,subfigure}
\usepackage{dcolumn}
\usepackage{bm}
\usepackage{float}
\usepackage{multirow}
\usepackage{cancel}
\usepackage{hyperref}
\usepackage{color}

\def\beq{\begin{equation}}
\def\eeq{\end{equation}}
\def\bea{\begin{eqnarray}}
\def\eea{\end{eqnarray}}

\def\3Eqs#1#2#3{Eqs.\ (\ref{#1}), (\ref{#2}) and (\ref{#3})}

\def\within#1to#2/{#1\mbox{ to }#2}
\def\lbullet#1,#2,#3,#4/{\Text(#1,#2)[c]{$\bullet$}\Line(#1,#2)(#3,#4)}
\def\textcite#1{Ref.~\cite{#1}}

\def\newu{$U(1)_{L_\mu-L_\tau}$}

\def\cens{CE$\nu$NS }

\def\hd0ino{\tilde{h^0_{d}}}

\def\hu0ino{\tilde{h^0_{u}}}

\def\e2spl{\ensuremath{^\clubsuit}}

\def\nn{\nonumber\\*}

\begin{document}
\hfill \preprint{ MI-TH-216}
\title{Probing $L_\mu-L_\tau$ models with CE$\nu$NS: A new look at the combined COHERENT CsI and Ar data}

\author{Heerak Banerjee}
\email{tphb@iacs.res.in}
\affiliation{School of Physical Sciences, Indian Association for the Cultivation of Science,
2A $\&$ 2B Raja S.C. Mullick Road, Kolkata 700 032, India}
\author{Bhaskar Dutta}
\email{dutta@physics.tamu.edu}
\affiliation{Mitchell Institute for Fundamental Physics and Astronomy,
Department of Physics and Astronomy, Texas A\&M University, College Station, Texas 77843, USA}
\author{Sourov Roy}
\email{tpsr@iacs.res.in}
\affiliation{School of Physical Sciences, Indian Association for the Cultivation of Science,
2A $\&$ 2B Raja S.C. Mullick Road, Kolkata 700 032, India}

\date{\today}

\begin{abstract}
The minimal gauged \newu model has long been known to be able to explain the tension between the theoretical and experimental values of the muon magnetic moment. It has been explored and tested extensively, pushing the viable parameter space into a very tight corner. Further, embedding the \newu model in a supersymmetric (SUSY) framework has been shown to relax some of these constraints and has recently been shown to explain the electron anomalous magnetic moment as well.
In this model, the logarithm of the mass ratio of third to second generation (s)leptons control the non-negligible kinetic mixing and may crucially alter many of the constraints. We confront both the non-SUSY and SUSY versions of this class of models with the CsI(2017), the recently released CENNS10 data from the liquid Argon detector as well as the updated CsI(2020) data of the COHERENT experiment. We use the recoil energy and timing binned data from CsI(2017) and the energy, time, and Pulse Shape Discriminator binned data from CENNS10 to find estimates for the model parameters
in a likelihood maximization test. We also show updated exclusions using all of the above data from the COHERENT Collaboration, as well as projected exclusions from the ongoing Coherent CAPTAIN-Mills experiment. The $(g-2)_\mu$ favored values of the \newu gauge coupling that are still unconstrained overlap with the estimates from COHERENT data within $1\sigma$. The combined COHERENT data is found to prefer the presence of the \newu gauge boson over the Standard Model at $\sim1.4\sigma$. 
The global minima of a chi-square deviation function using CsI(2020) as well as CENNS10 total counts has significant overlap with the $(g-2)_{\mu}$ favored parameter space in the context of the SUSY and non-SUSY $L_{\mu}-L_{\tau}$ models with a mediator mass in the $20-100$ MeV range.
\end{abstract}

\maketitle

\section{Introduction}

The prospect of observing coherent elastic neutrino-nucleus scattering (CE$\nu$NS) was first proposed roughly 47 years ago\cite{Freedman:1973yd}. However, the crucial discovery was of the fact that the scattering cross-section involved with the \cens process would be many times larger than the corresponding process with a single nucleon. The \cens cross-section was expected to scale approximately with the square of the neutron number of the nuclei\cite{Freedman:1973yd, Drukier:1983gj}.

In spite of the large predicted event rates, which meant that detectors with masses as small as a few kilograms could be used to detect it, \cens evaded observation for four decades, primarily because of the small (few keV) nuclear recoil energies involved. The COHERENT Collaboration finally reported its observation of the process with stopped-pion neutrinos on a CsI detector, 43 years after it was first proposed, and the recoil energy distribution agreed with the Standard Model (SM) within 1$\sigma$. The Collaboration also reported that the two-dimensional (energy and time) profile, fit to the maximum likelihood, prefers the presence of \cens against its absence at the 6.7$\sigma$ confidence level\cite{Akimov:2017ade}. Henceforth, we refer to this as the CsI(2017) dataset. This was followed up by the observation of \cens on a liquid Ar detector which also rejects the absence of \cens at $3.5\sigma$ with 3D-binned data in their recoil energy, timing, as well as pulse-shape-discriminator (PSD) data~\cite{Akimov:2020pdx}. The COHERENT Collaboration has also recently announced the impending publication of an update to the initial dataset from their CsI detector\cite{magnificient}, referred to as CsI(2020), which rejects the absence of \cens at $11.6\sigma$. Not only have these results added to the understanding of neutrino interactions and nuclear form factors\cite{Cadeddu:2018izq,Canas:2018rng,Cadeddu:2020lky,Cadeddu:2017etk,Papoulias:2019lfi,AristizabalSierra:2019zmy,Coloma:2020nhf} within the SM, but they have also provided a benchmark to test any propositions of physics beyond the SM that altered them.

These results have been used to test several new physics scenarios that result in neutrino nonstandard interactions (NSI)\cite{Barranco:2005yy,Scholberg:2005qs,Barranco:2007tz,Dutta:2015vwa,Lindner:2016wff,Dent:2016wcr,Coloma:2017egw,Shoemaker:2017lzs} and their electromagnetic properties\cite{Dodd:1991ni,Kosmas:2015sqa}. They have also been transcribed into constraining models involving sterile neutrinos\cite{Anderson:2012pn,Dutta:2015nlo,Kosmas:2017zbh}, neutrino magnetic moments\cite{Miranda:2019wdy,Cadeddu:2020lky}, light dark matter\cite{Dutta:2020vop,deNiverville:2015mwa} and light vector mediators of new physics\cite{Denton:2018xmq,Cadeddu:2020nbr,Dutta:2020enk,Amaral:2020tga,Miranda:2020zji}.
We build upon the method introduced in Ref.\cite{Dutta:2019eml} to compare the combined observations\cite{Akimov:2018vzs,Akimov:2020czh} of the COHERENT Collaboration from both CsI(2017) as well as the Ar (CENNS10) detectors 
\begin{figure*}[t]
    \centering
    \subfigure[\label{fig:csienergy}]{\includegraphics[width=8cm]{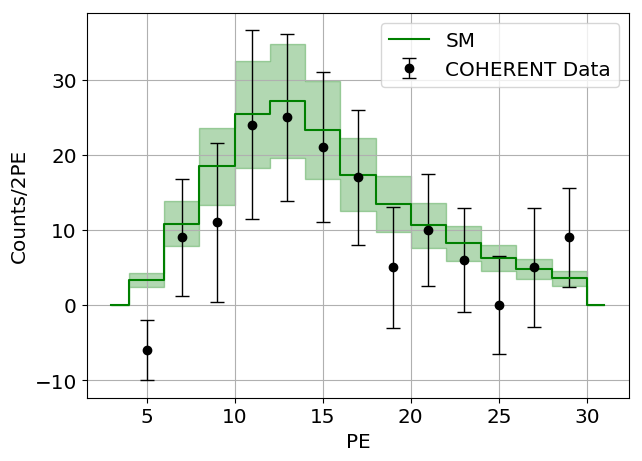}}
    \subfigure[\label{fig:csitiming}]{\includegraphics[width=8cm]{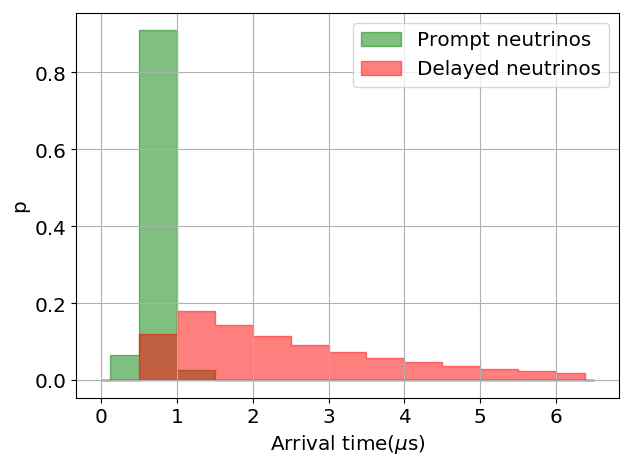}}
    \caption{The SM prediction as well as the observed \cens energy distribution from the CsI(2017) dataset of the COHERENT experiment is shown in Fig.\ref{fig:csienergy}~\cite{Akimov:2018vzs}. The \cens timing distribution generated from the event arrival time (the onset of scintillation) observation from the CsI(2017) dataset is shown in Fig.\ref{fig:csitiming}.}
    \label{fig:cevnscsi}
\end{figure*}
with the \cens prediction in both nonsupersymmetric (non-SUSY) and supersymmetric (SUSY) $L_\mu-L_\tau$ models in a likelihood maximization test. We also use the announced total count from the CsI(2020) data in conjunction with that from the Ar detector in a chi-square deviation test of the model.

The $L_\mu-L_\tau$ model as a viable extension of the SM has been studied in much detail\cite{He:1990pn,He:1991qd}. It has been shown to be able to successfully explain the muon 
magnetic moment \cite{Baek:2001kca,Ma:2001md} that has been observed to be at $\sim 4.2\sigma$\cite{Bennett:2006fi,Albahri:2021kmg}($\delta a_\mu = 2.51(59)\times 10^{-9}$) from its SM prediction\cite{Aoyama:2012wk,Aoyama:2019ryr,Czarnecki:2002nt,Gnendiger:2013pva,Davier:2017zfy,Keshavarzi:2018mgv,Colangelo:2018mtw,Hoferichter:2019gzf,Davier:2019can,Keshavarzi:2019abf,Kurz:2014wya,Melnikov:2003xd,Masjuan:2017tvw,Colangelo:2017fiz,Hoferichter:2018kwz,Gerardin:2019vio,Bijnens:2019ghy,Colangelo:2019uex,Blum:2019ugy,Colangelo:2014qya}. It has also been used to explain neutrino masses and mixing \cite{Ma:2001md,Heeck:2011wj,Baek:2015mna,Biswas:2016yan}, 
dark matter \cite{Altmannshofer:2016jzy,Biswas:2016yan,Biswas:2016yjr,Patra:2016shz,Biswas:2017ait,Arcadi:2018tly,Kamada:2018zxi,Foldenauer:2018zrz}, 
Higgs boson flavor violating decays \cite{Crivellin:2015mga,Altmannshofer:2016oaq}, and B-decay 
anomalies \cite{Altmannshofer:2014cfa,Crivellin:2015mga,Altmannshofer:2016jzy,Ko:2017yrd,Baek:2017sew} for different choices of the parameter space that have some intersection amongst themselves. The extension was embedded in a SUSY framework and the associated phenomenology was discussed with respect to $(g-2)_\mu$, neutrino masses and mixing and charged lepton flavor violation in Ref.\cite{Banerjee:2018eaf} and dark matter in Ref.\cite{Das:2013jca}. A more recent study of the SUSY $L_\mu-L_\tau$ model shows that both electron and muon anomalous magnetic moments may be explained simultaneously~\cite{Banerjee:2020zvi}. 

The modification to the SM \cens rate due to the $L_\mu-L_\tau$ gauge boson proceeds via the kinetic mixing ($\epsilon$), a necessarily small parameter ($\lesssim 10^{-5}$). As a consequence, the modification is dominated by the interference term, which is destructive as long as the sum of the u- and d-quark beyond the Standard Model (BSM) couplings to the neutrinos remain positive. This holds for all but a few pathological parameter choices for the \newu model. This is owing to the fact that these couplings in the $L_\mu-L_\tau$ model are proportional to $\epsilon$Q$_{em}$. While the presence of the $L_\mu-L_\tau$ gauge boson thus serves to reduce the \cens rate, the CsI detector has observed a count lower than that predicted by the SM and the Ar detector has reported an opposite trend. It is true that both results are within $1\sigma$ of the SM prediction, still, a more rigorous analysis involving the full energy as well as timing data has shown that the CsI data prefers new physics scenarios that lead to a reduction of the number of counts at $\sim2\sigma$\cite{Dutta:2019eml}. We re-evaluate the situation in the light of the recent Ar data adding a pull in the opposite direction. 

Our analysis finds that although the preference for destructive BSM effects has a marked reduction, it still remains slightly larger than $1\sigma$. We use our analysis to show updated exclusions on the $L_\mu-L_\tau$ parameter space. The exclusions from the Ar data are much stronger than the corresponding exclusions from CsI(2017), however the exclusions from the combined data adhere closely to those from the CsI(2017) data. The CsI(2020) combined with the Ar data lays much stronger bounds than the former, still; they are not the strongest constraints in the relevant parameter space. However, the preferred region of parameter space from the likelihood test, as well as the global minima of the chi-square deviation has significant overlap with the $(g-2)_{\mu}$ preferred region in the context of the $L_\mu-L_\tau$ model. In addition, we also show projected exclusions from the Coherent CAPTAIN-Mills (CCM)\cite{ccm1,ccm2} experiment which is another experiment looking for \cens with stopped-pion neutrinos on a liquid Ar detector. In comparison with the COHERENT experiment, CCM uses a less intense neutrino beam but has a much larger detector fiducial volume ($\sim$5 tons as opposed to $\sim$24 kgs at COHERENT). In case the CCM observations correspond to the SM predictions, the exclusions would be the strongest to date in the pertinent region of the parameter space. 

\section{CE$\nu$NS at COHERENT}
\begin{figure*}[t]
    \centering
    \subfigure[\label{fig:larenergy}]{\includegraphics[width=5.8cm]{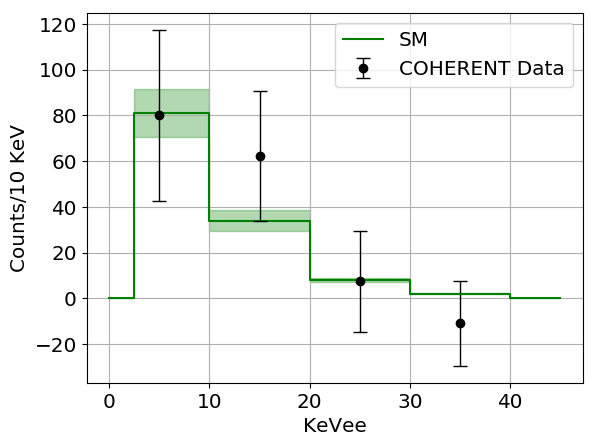}}
    \subfigure[\label{fig:lartiming}]{\includegraphics[width=5.8cm]{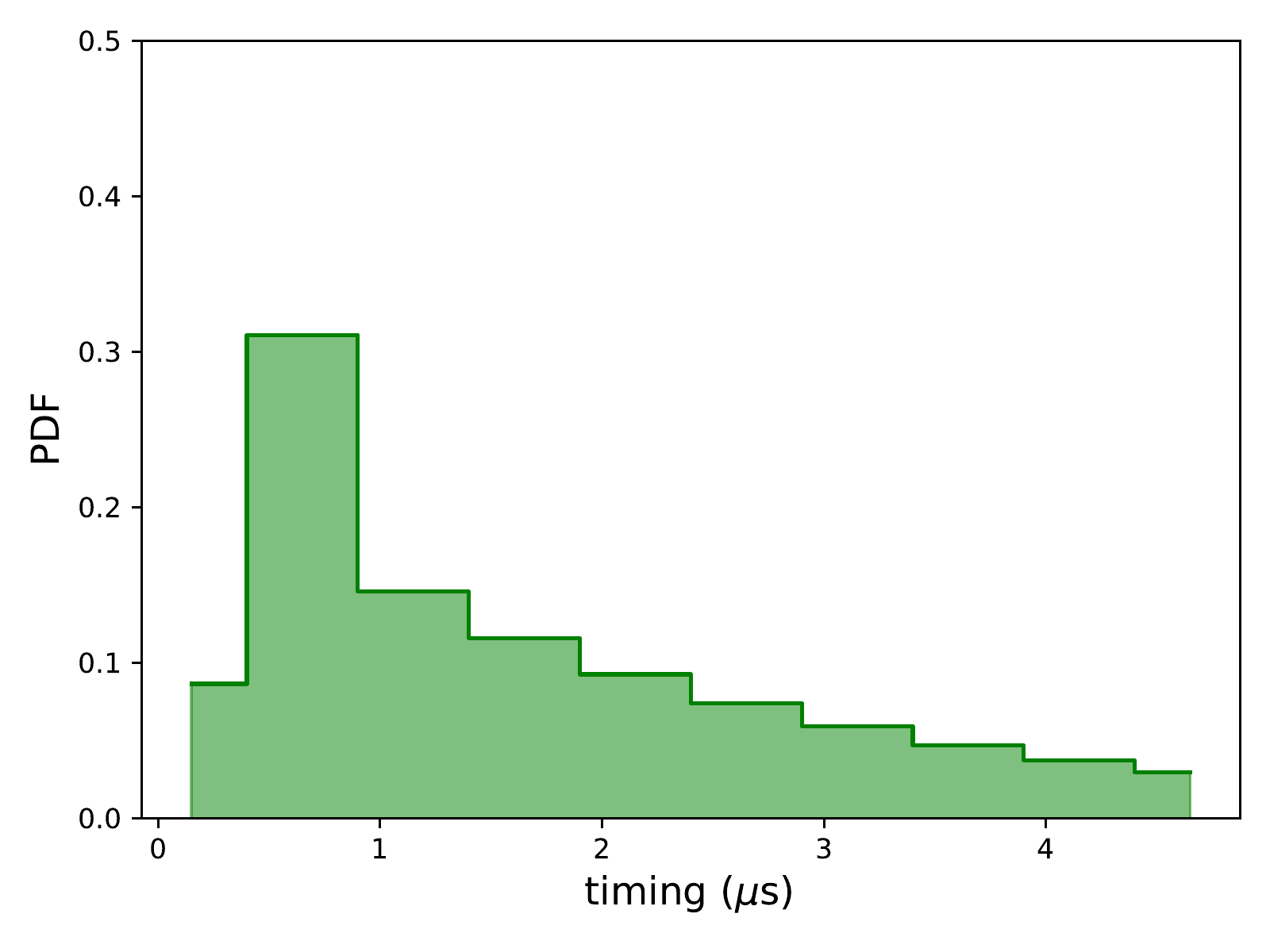}}
    \subfigure[\label{fig:larf90}]{\includegraphics[width=5.8cm]{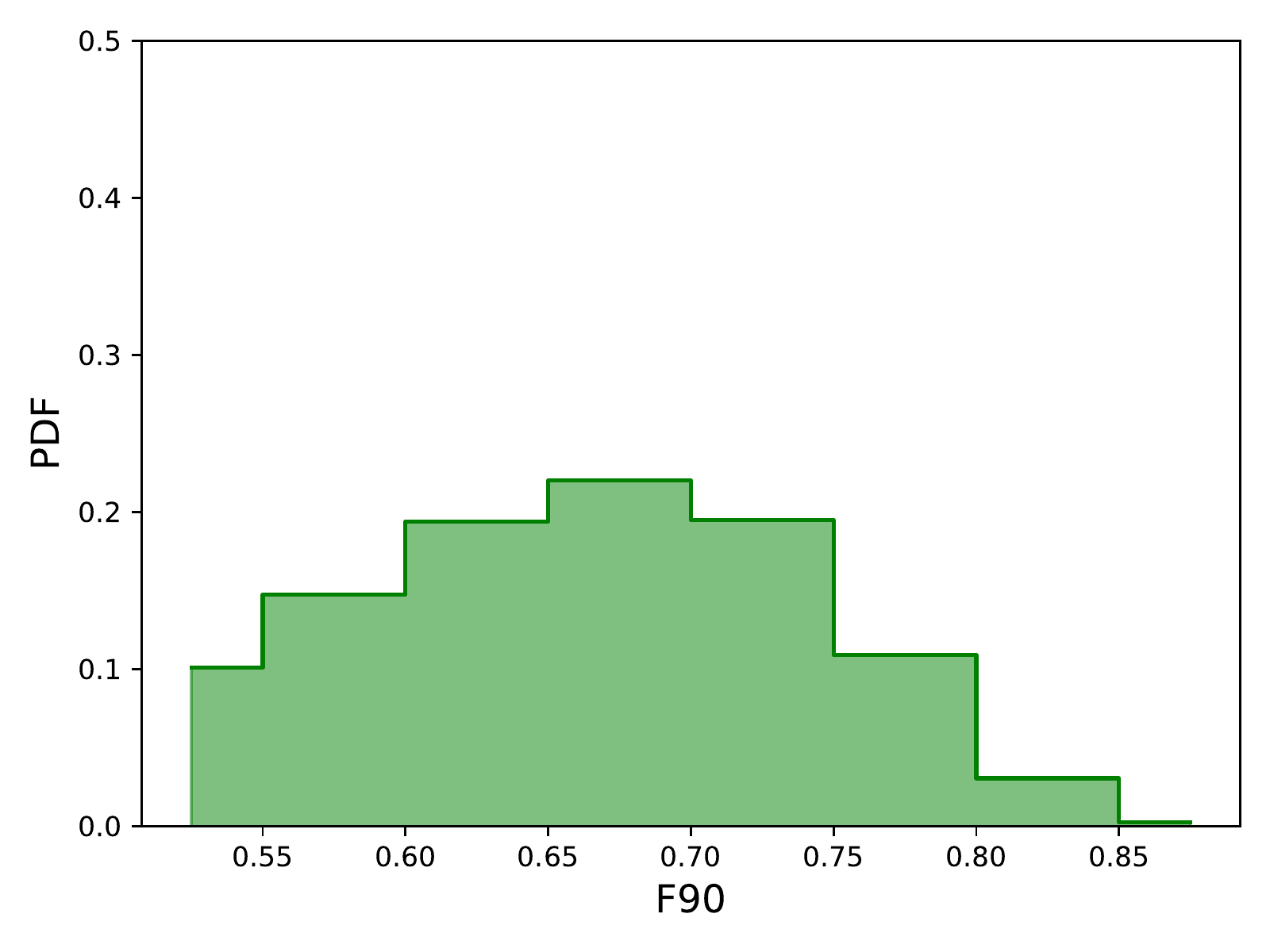}}
    \caption{\cens energy (\ref{fig:larenergy}), timing (\ref{fig:lartiming}) and F90 (\ref{fig:larf90}) distributions from the CENNS10 detector~\cite{Akimov:2020pdx}. Also shown in Fig.\ref{fig:larenergy} is our SM \cens prediction for CENNS10 along with the associated total uncertainty shown shaded green.}
    \label{fig:cevnslar}
\end{figure*}
The COHERENT Collaboration was designed to detect and study \cens over a range of detectors. It uses a high intensity neutrino beam produced at the Spallation Neutron Source (SNS) of the Oak Ridge National Laboratory. A 1 GeV proton beam incident upon a mercury target at 60 Hz produces $\sim0.08-0.09$ neutrinos per flavor per proton on target (POT). Upon hitting the Hg target, the proton beam produces pions. 

The negative pions are captured almost entirely within the target while the positive pions decay to produce the neutrinos,
\begin{align}
&\pi^+ \rightarrow \mu^+ {\nu}_\mu&\qquad &{\rm (Prompt)}\nn
&\mu^+ \rightarrow e^+ \nu_e \bar{\nu}_\mu& \qquad &{\rm (Delayed)}.
\end{align}
The monoenergetic $\nu_\mu$ from the first pion decay are ``Prompt" and arrive at the target within a very short span of time ($\lesssim 1.5\mu s$) after the POT. The $\nu_e$ and $\bar{\nu}_\mu$ from the subsequent muon decay are ``Delayed" and have an energy profile as they are produced in a three-body decay. The neutrino flux from the SNS may be described by,
\begin{eqnarray}
\frac{{\rm d}\Phi_{\nu_\mu}}{{\rm d}E_\nu} &=& \eta \delta\left[E_\nu - \frac{m_\pi^2 - m_\mu^2}{2m_\pi}\right]\nn
\frac{{\rm d}\Phi_{\bar{\nu}_\mu}}{{\rm d}E_\nu} &=& \eta \frac{64E_\nu^2}{m_\mu^3}\left(\frac34 - \frac{E_\nu}{m_\mu}\right)\nn
\frac{{\rm d}\Phi_{\nu_e}}{{\rm d}E_\nu} &=& \eta \frac{192E_\nu^2}{m_\mu^3}\left(\frac12 - \frac{E_\nu}{m_\mu}\right)\nn
\eta &=& \frac{rN_{\rm POT}}{4\pi L^2},
\end{eqnarray}
where $r=$ the number of neutrinos per flavor produced per proton on target, $N_{\rm POT}$= the total number of protons on target during the data-taking period, and $L=$ the distance of the detector from the neutrino source.
This results in a distinct timing spectra for the incoming neutrinos at the SNS as shown in Fig.\ref{fig:csitiming}.

Data from the observation of \cens at two different detectors has been made public by the COHERENT Collaboration to date: the CsI and the CENNS10 liquid Ar detectors.

\subsection{Backgrounds}
\begin{figure*}[t]
    \centering
    \subfigure[\label{fig:cevnsdiag}]{\includegraphics[width=6cm]{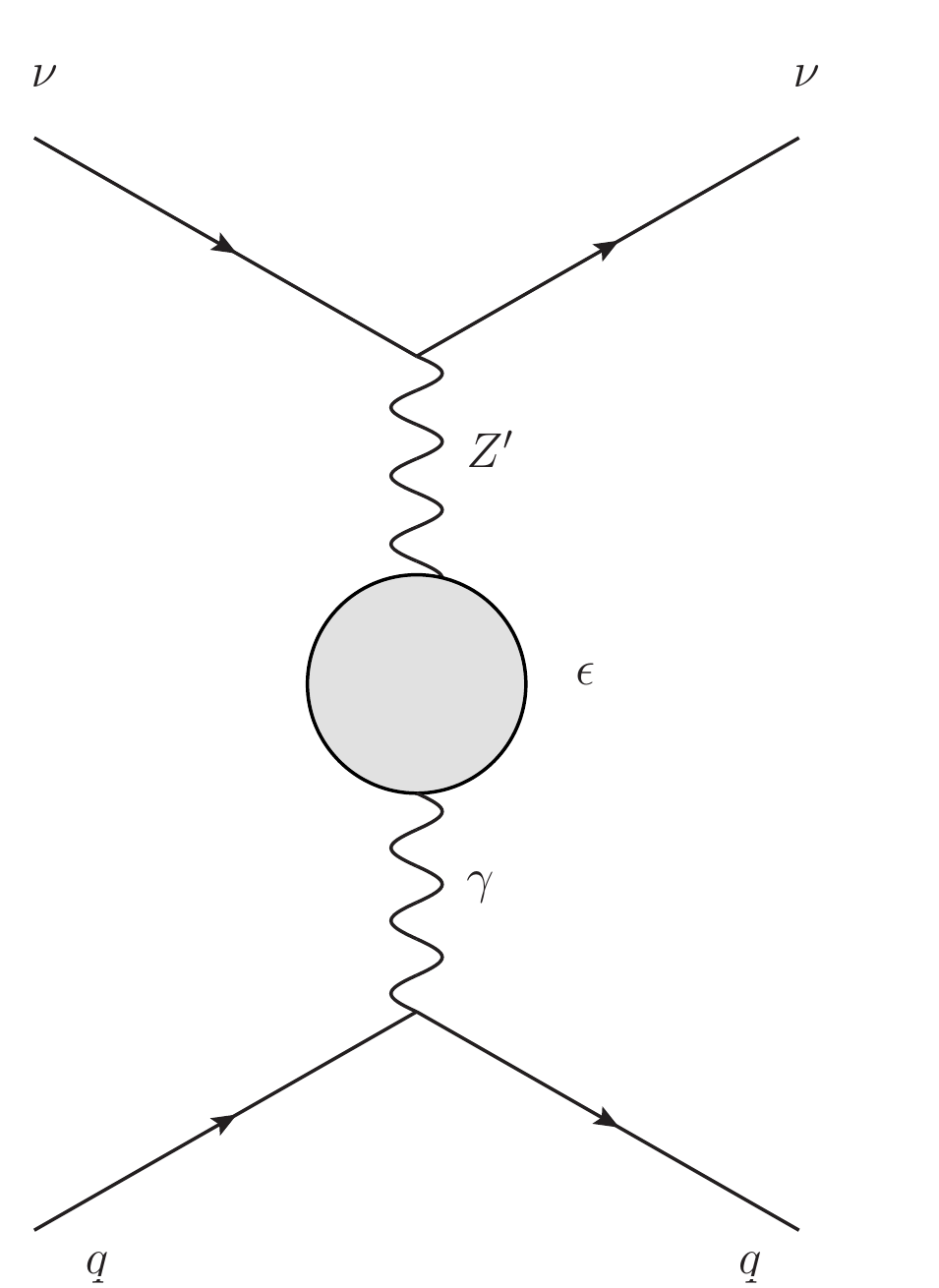}}
    \centering
    \subfigure[\label{fig:cstest}]{\includegraphics[width=8cm]{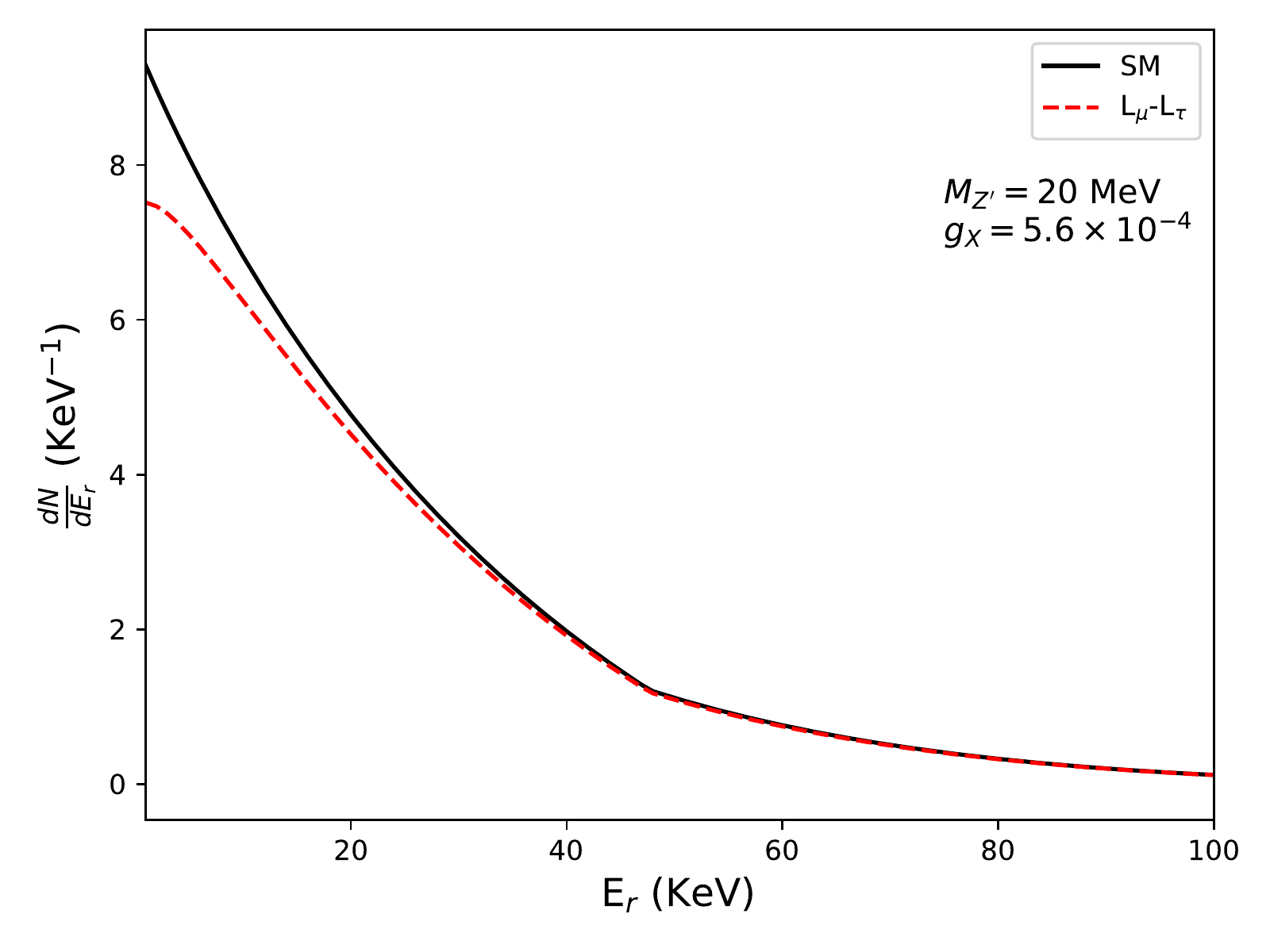}}
    \caption{Fig.\ref{fig:cevnsdiag} shows the diagram by which an additional gauge boson, $Z^\prime$, modifies the SM \cens rate. Fig.\ref{fig:cstest} shows the variation of the differential \cens event rate at the COHERENT CENNS10 detector with nuclear recoil energy. The black line shows the SM prediction while the red line corresponds to the modified rate in $L_\mu-L_\tau$ for a 20 MeV $Z^\prime$ with a gauge coupling of $5.6\times 10^{-4}$ and a vanishing SUSY contribution to $\epsilon$.}
    \label{fig:cevns}
\end{figure*}
For both detectors, there are two broad classes of backgrounds that affect the analysis. The first, steady state backgrounds, are present in both beam-on and beam-off data and may be handled using prescribed methods of background subtraction as described in Refs.\cite{Akimov:2018vzs,Akimov:2020pdx}. The second, beam-related-neutrons (BRN) and neutrino-induced-neutrons (NIN) occur in time with the SNS beam and are hence more difficult to separate. The BRN can be further classified into prompt and delayed BRN, the former are the fast neutrons produced by the SNS target itself while the latter are produced by the neutrinos interacting with the detector shielding. Both of these backgrounds generate events similar in nature to the \cens events and need to be correctly accounted for. The COHERENT Collaboration uses the different recoil energy and timing distributions for these events to discriminate between this background and the \cens signal.

The COHERENT Collaboration has released data from dedicated measurements of both the kinds of backgrounds for CsI as well as CENNS10 detectors. In our analyses we compare the total observed counts with the model prediction added to these background counts using appropriate uncertainties for all the distributions. This is consistent with the method used by the Collaboration itself to generate the best-fit \cens and background counts. As a cross-check, we use our method to re-evaluate these best-fit counts and find a very close match with those reported by the Collaboration.

\subsection{CsI}\label{csi}

This was the first \cens detector installed at the SNS and it was composed of a 14.6Kg CsI[Na] crystal. It was installed at a distance of 19.3 m from the neutrino source at $90^{\circ}$ to the proton beam direction. It could detect nuclear recoil energies as small as $4.25$ keV although the efficiency of detection and background distinction was poor below $\sim 10$ keV. It was in place from 2015 to 2019 and was used to make the first observation of \cens with 171.7 beam-on live-days, which amounted to approximately $1.76\times 10^{23}$ protons on target.
The quenching factor for the CsI detector,
\begin{equation}
    n_{\rm PE} = 1.17\left(\frac{E_r}{\rm keVnr}\right)\label{eqn:csiqf}
\end{equation}
describes the number of photoelectrons detected by the photo multiplier tubes per keV nuclear recoil energy produced in a \cens event. Reference \cite{Akimov:2018vzs} describes the detector acceptance efficiency in terms of the detected number of photoelectrons (x) by the function
\begin{eqnarray}
f(x) = \frac{a}{1 + exp(-k(x-x_0))}\Theta(x-5) \label{eqn:acceptance}
\end{eqnarray}
where,
\begin{align*}
a &= 0.6655^{+0.0212}_{-0.0384} \\
k&=0.4942^{+0.0335}_{-0.0131}\\
x_0&=10.8507^{+0.1838}_{-0.3995}.
\end{align*}
Once the number of photoelectrons is mapped to nuclear recoil energies using Eq.\ref{eqn:csiqf}, we get the acceptance $\mathcal{A}(E_r)$.
The CsI detector reported~\cite{Akimov:2017ade} an observation of $134\pm22$ \cens events at a $6.7\sigma$ C.L. against an SM prediction of $173\pm48$ events in 2017. The energy, as well as the timing distribution [the probability distribution function (PDF) generated from observation] of the observed data, as reported, is shown in Fig.\ref{fig:cevnscsi}. This observation has a total fractional uncertainty of $28\%$ which is a combination of neutrino flux ($10\%$), nuclear form factor ($5\%$), quenching factor ($25\%$), and cut-acceptance ($5\%$) uncertainties. The COHERENT Collaboration has recently announced the publication of additional data from the CsI detector\cite{magnificient}, however only the total \cens and background counts have been made public along with the respective uncertainties. We use the public data from the first COHERENT data release in the likelihood based analysis (discussed in Sec.\ref{sec:lkltest}) and the recently announced total counts in a $\chi^2$ test.

\subsection{CENNS10}\label{ar}

This is a liquid Argon detector with 24Kg of active fiducial mass placed 27.5m away from the SNS target off the proton beam axis by $\sim 135^\circ$ and operational since 2017. With a threshold nuclear recoil energy of $\sim 3$ keV, the detector recorded data worth approximately $1.38\times 10^{23}$ POT. Reference \cite{Akimov:2020pdx} describes the quenching factor,
\begin{equation}
    E_{ee} = (0.246 + (7.8\times10^{-4}/keV_{nr})E_{nr}) \times E_{nr}.
\end{equation}
The acceptance, $\mathcal{A}(E_r)$, for the CENNS10 detector is interpolated from efficiency data binned in nuclear recoil energy provided with the data release. In addition to the energy and timing information of the observation, this data release also includes the PSD distribution. This observable is used to distinguish between electron and nuclear recoils in Ar. More on the PSD variable (F90) and the cuts used on it can be found in Ref.\cite{Akimov:2020pdx}. We use F90 cuts identical to Analysis A described in this reference.

The CENNS10 detector reported~\cite{Akimov:2020pdx} the observation of $159\pm43({\rm stat})\pm 14({\rm syst})$ \cens events at a significance of $3.5\sigma$, as opposed to a SM prediction of $128\pm17$ events. The energy, timing and F90 distributions from the COHERENT Collaboration are shown in Fig.\ref{fig:cevnslar}. The involved uncertainties were divided into two part. The first set combines the quenching factor ($1\%$), calibration ($0.8\%$), detector efficiency ($3.6\%$), prompt light fraction ($7.8\%$), form factor ($2\%$), and neutrino flux ($10\%$) uncertainties for a total of $13\%$. This is an uncertainty on the \cens rate prediction and is called the prediction uncertainty for the rest of this work. The second set includes uncertainties in the F90 energy dependence($4.5\%$), neutrino arrival time($2.7\%$), BRN energy shape($5.8\%$), BRN arrival time-mean($1.3\%$), and BRN arrival time-width($3.1\%$) and leads to a total uncertainty of $8.5\%$ on the \cens best-fit value. This set will be referred to as the fit uncertainty. Public data from the CENNS10 detector is used in a likelihood as well as an $\chi^2$ test in conjunction with the observations from the CsI detector.

\subsection{Update to CsI data}\label{sec:csi-update}

The COHERENT Collaboration has recently announced the publication of additional data from its CsI detector~\cite{magnificient} that has observed $306\pm 20$ events against an SM prediction of $333\pm11({\rm th})\pm42({\rm ex})$. While the full details of the observation have not been made public yet, we have enough information to conduct a preliminary analysis with it. The Collaboration has updated its signal prediction as well as the \cens data analysis which has lead to a significant reduction in the overall uncertainty of the signal from $28\%$ to $13\%$. This has also resulted in a reduction in both the predicted and observed \cens count.

The update releases data taken on the CsI[Na] detector until June 2019, which amounts to $\sim 2.2\times$ more protons on target compared to the previous CsI data release. In the absence of detailed quenching factor and efficiency information from the new analysis, we use those from the previous data release. According to the Collaboration, the new analysis leads to a $\sim 9\%$ reduction in the SM prediction for the CsI(2017) dataset. We find that the predicted signal for the CsI(2020) dataset using the earlier methods requires $\sim 14\%$ reduction to match the new SM prediction released by the Collaboration. We also have information on the steady  state background ($1273\pm 38$), BRN ($17\pm4$), and NIN ($5\pm2$), and the corresponding uncertainties for the new data.

Since we have only the total observed counts for both the signal and background in the new dataset, adopting a full likelihood maximization routine like the ones for the CENNS10 and old CsI data is overkill. Instead, we use a chi-square fit of the model parameters to the observed data in order to get the updated exclusions on the $M_{Z^\prime}-g_X$ parameter space,
\begin{equation}
    \chi^2_{\rm CsI}= {\rm Min}_\alpha \left[\frac{(N^{obs} - (1 + \alpha)N^{th})^2}{\sigma_{obs}^2} + \left(\frac{\alpha}{\sigma_{\alpha}}\right)^2\right]\label{eqn:chi-csi},
\end{equation}
with $N^{obs}=306$ denoting the total observed \cens count and $N^{th}$ denoting the model prediction. The quantities $\sigma_{obs}^2 =(\delta N^{obs})^2 + (\delta N^{SS})^2 + (\delta N^{BRN})^2 + (\delta N^{NIN})^2$ and $\sigma_{\alpha}=0.13$ parametrize the total uncertainties on the observed and predicted counts, respectively, 
\begin{align}
    \delta N^{obs} = 20 \nonumber\\
    \delta N^{SS} = 38 \nn
    \delta N^{BRN} = 4 \nn
    \delta N^{NIN} = 2.
\end{align}

In order to combine the CENNS10 data with this, we must define a similar chi-square function for it,
\begin{equation}
    \chi^2_{\rm Ar}= {\rm Min}_\beta \left[\frac{(N^{obs} - (1 + \beta)N^{th})^2}{\sigma_{obs}^2} + \left(\frac{\beta}{\sigma_{\beta}}\right)^2\right]\label{eqn:chi-ar}.
\end{equation}
with $N^{obs}=159$ and $N^{th}$ denoting the model prediction for the CENNS10 analysis. The quantities $\sigma_{obs}^2 =(\delta N^{obs})^2 + (\delta N^{SS})^2 + (\delta N^{pBRN})^2 + (\delta N^{dBRN})^2$ and $\sigma_{\beta}=0.1328$ are the total uncertainties on the observed and predicted counts as for the previous case,
\begin{align}
    \delta N^{obs} = 43 \nonumber\\
    \delta N^{SS} = 34 \nn
    \delta N^{pBRN} = 23 \nn
    \delta N^{dBRN} = 11.
\end{align}
In our analysis we define the total chi-square deviation as the sum of the two individual ones and minimize the chi-square deviation over the $\alpha$ and $\beta$ nuisance parameters.

\section{\cens at CCM}

The proposed CCM experiment is designed to detect \cens on a liquid Argon detector with a $\sim 5$ ton fiducial mass using neutrinos from the Lujan target at the Los Alamos National Laboratory~\cite{ccm1,ccm2,CCM:2021leg}. The detector is placed 20 m from the Lujan target that produces $\sim 0.04$ neutrinos per flavor per POT. The Lujan facility uses an 800 MeV proton beam impinging upon a tungsten target at 20 Hz to produce pions that decay and produce neutrinos with the same energy shape as those at the SNS. 

We assume 5000 hours of operation per year, for three years, for our analysis which is approximately $3\times 10^{22}$ POT for the entire period. In the absence of any quenching factor, efficiency, or background information we only use a detector threshold of 25 keV and the \cens count alone for evaluating the projected exclusions. The detector efficiency above the threshold was kept at $100\%$ and a 1 PE/keVnr quenching factor was used for our analysis. This was done for the sake of easy scalability once the full data from the detector becomes available. 

\section{Signal prediction}\label{sec:sigpred}

In this section we shall describe the \cens cross section as it arises in the SM, discuss how it may be modified in the $L_\mu-L_\tau$ model (in both the presence and absence of SUSY), and describe how we generate the signal prediction for the experiments.

In coherent neutrino-nucleus scattering events, the momentum transfer between the neutrino and the nucleus is smaller than or comparable to inverse the size of the nucleus [$E_r \lesssim\mathcal{O}(50)$ MeV]. The axial vector contribution to this process is proportional to the nucleus spin~\cite{Barranco:2005yy}, and hence is negligible in comparison to the vector contribution.
The cross section for \cens scattering on a nucleus of mass $m_N$, atomic number $Z$, and neutron number $N$ is
\begin{equation}
	\dv{\sigma}{E_r} = \frac{G_F^2Q_V^2}{2\pi}m_N\left(1-\left(\frac{m_NE_r}{E_{\nu}^2}\right) + \left(1-\frac{E_r}{E_{\nu}}\right)^2\right)\vert F(q)\vert^2
	\label{eqn:dsigder}
\end{equation}
where $F\left(q\right)$ is the nuclear form factor and $Q_V = Z\left(2g^u_v+g^d_v\right)+N\left(g^u_v+2g^d_v\right)$. We use $g_v= T_3-2Q_{\rm em}{\rm{sin}}^2\theta_W$, the quark vector coupling to the $Z$-boson where $T_3$ is the third component of the weak isospin, $Q_{\rm em}$ is the electromagnetic charge, and $\theta_W$ is the weak mixing angle.

The form factor $F\left(q\right)$ is a description of the nucleon distribution within the nucleus and follows from its Fourier transform. It is captured by several parametrizations, e.g., the Helm~\cite{Helm:1956zz}, Symmetrized Fermi~\cite{Sprung1997}, and Klein Nystrand~\cite{Klein:1999qj} approaches. Sticking to the conventions adopted by the COHERENT Collaboration in their own analysis of CsI and CENNS10 data, we use the Helm and Klein-Nystrand parametrizations for the two detectors, respectively. Note, however, that the choice of form factor parametrization has little or no effect on the results obtained.  

The Helm form factor can be described by
\begin{equation}
    F_H(q^2) = 3 \frac{j_1(qR_0)}{qR_0}e^{-q^2s^2/2}
\end{equation}
where q=$\sqrt{2m_NE_r}$ is the absolute value of exchanged three-momentum, s=$0.9$ fm is the nuclear skin width and $j_1(x)=\sin{x}/x^2 - \cos{x}/x$ is the spherical Bessel function of order one. The neutron charge radius is given by
\begin{equation}
    R_n^2 = \frac35 R_0^2 + 3s^2.
\end{equation}
Similarly, the Klein-Nystrand form factor is described by,
\begin{equation}
    F_{KN}(q^2)= 3 \frac{j_1(qR_0)}{qR_0}\frac{1}{1 + q^2 a_k^2}.
\end{equation}
The parameter $a_K=0.7$ fm enters the expression for the neutron charge radius,
\begin{equation}
     R_n^2 = \frac35 R_0^2 + 6a_k^2.
\end{equation}
The size of the neutron distribution radius $R_n$ is taken to be 4.7 fm and 4.1 fm for the analyses involving CsI and Ar, respectively~\cite{Cadeddu:2020lky}.

The SM \cens cross section may be modified in the presence of an additional gauge boson coupling to the neutrinos and/or quarks. The $L_\mu-L_\tau$ gauge boson has direct couplings to second- and third-generation leptons as well as their scalar superpartners in the presence of SUSY. It also couples to any other charged field e.g. the quarks, via its kinetic mixing with the photon. It has the following interactions with the neutrinos and the quarks:
\begin{equation}
\mathcal{L} \supset Z^\prime_\mu\left(g_{x}\,Q_X^i\,\bar{\nu}_i\gamma^{\mu}\nu_i +  \epsilon\, e\, Q_{\rm em}\,\bar{q}\gamma^{\mu}q\right).
\label{Zprime}
\end{equation}
Here $Q_X$ are the charges of the neutrinos under the additional gauged symmetry as defined in Ref.\cite{Banerjee:2020zvi}, and $\epsilon$ is the kinetic mixing parameter. It modifies the \cens cross section via the kinetic mixing of the $Z^\prime$ with the photon [see Fig.\ref{fig:cevnsdiag}]. In order to implement the modification of the SM \cens rate by the additional contributions it is enough to make the replacement
\begin{equation}
g_v \,\rightarrow\, g_v
+ \frac{g_{x}\, \epsilon\, e\, Q_{\rm em}}{2\sqrt{2}G_F\, (q^2 + M_{Z'}^2)},
\label{eq:SMmod}
\end{equation}
in Eq. \ref{eqn:dsigder}. This replacement modifies the quantity $Q_V$ in Eq. \ref{eqn:dsigder} which may be captured by the transformation
\begin{equation*}
Q_V \rightarrow Q_V^{SM} + Q_V^{BSM}
\end{equation*}
where 
\begin{align}
Q_V^{BSM} &= (A+Z)\frac{g_{x}\, \epsilon\, e\, Q^u_{\rm em}}{2\sqrt{2}G_F\, (q^2 + M_{Z'}^2)}\nn 
&+ (A+N) \frac{g_{x}\, \epsilon\, e\, Q^d_{\rm em}}{2\sqrt{2}G_F\, (q^2 + M_{Z'}^2)}\label{eqn:qvmod}
\end{align}
and A, Z, and N are the mass number, atomic number, and neutron number, respectively, for the target nucleus (Cs, I, or Ar in our case, depending on the detector).
The two terms in Eq.\ref{eqn:qvmod} represent the u-quark and d-quark couplings to the neutrinos multiplied by (A+Z) and (A+N) respectively and differ only by their electromagnetic charge. Note that in our case, owing to the inherent smallness of $\epsilon$ and hence $Q_V^{BSM}$, the interference term $2Q_V^{SM}Q_V^{BSM}$ in the differential cross-section dominates over the pure BSM part of it. Moreover, since $Q_V^{SM}\sim -0.5 N$, the interference is always destructive as long as the right hand side of eqn.\ref{eqn:qvmod} remains positive. This holds for the $L_\mu-L_\tau$ model except for a few pathological parameter choices since the u-$\nu$ BSM coupling is positive and twice the d-$\nu$ coupling, which is negative. Thus, the \cens count suffers a reduction in this model.

Figure \ref{fig:cstest} illustrates this effect for a choice of the $L_\mu-L_\tau$ parameters. The black solid line in this figure denotes the SM prediction of the differential \cens event rate at the CENNS10 detector of the COHERENT experiment. The red dashed line shows the same in the presence of a 20 MeV $L_\mu-L_\tau$ $Z^\prime$ with the gauge coupling ($g_x$) set to $5.6\times 10^{-4}$ and the SUSY contribution to $\epsilon$ taken to be vanishing. This choice of parameters can explain $(g-2)_{\mu, e}$ in the presence of SUSY~\cite{Banerjee:2020zvi} and does not predict an observable charged lepton flavor-violating decay width, even when neutrino masses and mixing may be explained in conjunction~\cite{Banerjee:2018eaf}.

The presence of kinetic mixing that mediates the $Z^\prime$ contribution to \cens may be assumed to be vanishing at the tree level, but it is unavoidably generated radiatively. In particular, the one-loop kinetic mixing in this class of models may be non-negligible in spite of super-heavy BSM fields entering the loops~\cite{Banerjee:2018mnw},
\bea
\epsilon &=& \frac{8 e g_X}{(4 \pi)^2} \int_0^1 x(1-x)\ln\frac{m^2_\tau - x(1-x)q^2}{m^2_\mu - x(1-x)q^2}dx \nn
&+&\frac{2 e g_X}{(4 \pi)^2} \int_0^1 (1-2x)^2 \ln\frac{m^2_{\tilde \tau} - x(1-x)q^2}{m^2_{\tilde \mu} - x(1-x)q^2}dx.
\label{eqn:km}
\eea 
This expression for $\epsilon$ involves both charged leptons and their scalar superpartners. As discussed in Ref.\cite{Banerjee:2018mnw}, this quantity stands out in its dependence on only the ratio of the (s)lepton masses. While the ratio of the second and third generation leptons are fixed, we take the ratio $r=m_{\tilde \tau}/m_{\tilde \mu}$ as an additional model parameter. In the absence of SUSY, only the first term contributes, as would be the case for degenerate sleptons ($r=1$) even in its presence. It also assumes left-right degeneracy amongst the charged sleptons for the sake of simplicity. Relaxing these assumptions would only lead to the dependence of $\epsilon$ on additional mass ratios beyond $r$. We use Eq.\ref{eqn:km} to provide the value of $\epsilon$ in the replacement defined in Eq.\ref{eq:SMmod} in order to generate the modified differential cross section. For both Eqs.\ref{eq:SMmod} and \ref{eqn:km}, the transferred momentum is $q^2= 2 M_N E_r$, where $M_N$ is the nuclear mass and $E_r$ is the nuclear recoil energy. For the mediator mass ($M_{Z^\prime}$) regime that is pertinent for the $L_\mu-L_\tau$ model ($\lesssim$ 1 GeV), $q^2$ is comparable to $M_{Z^\prime}^2$. Unlike general neutrino NSIs, the BSM modification to the \cens rate is thus nuclear recoil energy dependent. The signal shape in the $E_r$ space due to this variation is taken into account by our likelihood function in the ensuing comparison. For a study of the impact of \cens observations on neutrino NSI parameter space see, for example, Ref. \cite{Dutta:2020che}. 

From the differential cross section, the total event rate in each recoil energy bin may be evaluated using,
\begin{equation}
    N_{CE\nu NS} = N_D\, \int {\rm d}E_r\int {\rm d}E_\nu \, \dv{\Phi_\nu}{E_\nu}\, \dv{\sigma}{E_r}\, \mathcal{A}(E_r)
\end{equation}
where $N_D$ represents the number of target nuclei in the detector mass and is given by 
\begin{equation}
    N_D = g_{\rm mol}\, \frac{m_{\rm det}\, N_A}{m_{\rm mol}},
\end{equation}
where, $m_{\rm det}$ is the detector mass, $N_A$ is the Avogadro's number, $m_{\rm mol}$ is the molar mass of the detector molecule and $g_{\rm mol}$ is the number of atoms in a single detector molecule.

The presence of beyond-SM coupling of the neutrinos to quarks modifies the SM \cens rate. For opposite sign couplings of the neutrinos to up- and down- type quarks, as is the case in the presence of an $L_\mu-L_\tau$ gauge boson, this effect is destructive as long as the two couplings are similar in magnitude.
As a result, the presence of the $L_\mu-L_\tau$ gauge boson serves to reduce the SM \cens rate for much of the parameter space, especially for low nuclear recoil energies [see Fig.\ref{fig:cstest}].  Once the nuclear recoil energy spectrum has been generated, we use the timing and F90 distributions shown in Figs.\ref{fig:csitiming}, \ref{fig:lartiming}, and \ref{fig:larf90} to generate the full experimental prediction for the respective detectors.  

\section{Likelihood test}\label{sec:lkltest}

This section describes the statistical analysis comparing the model predictions with experimental data from the COHERENT Collaboration. We adopt the method described in Ref.\cite{Dutta:2019eml} for the analysis of the CsI data using both energy and time binning. We build upon this to devise a method to test our model against the full 3D-binned CENNS10 data as well, the goal being an analysis using combined CsI and CENNS10 data binned in recoil energy and timing as well as PSD observables.

While defining the likelihood function for the CsI as well as CENNS10 data, we assume that the total count (\cens + background) follows a Poisson distribution in each bin. The model prediction added to the observed background counts acts as the parameter for the Poisson distributions used. The likelihood function is marginalized over the uncertainties in each of the separate distributions assuming a Gaussian envelope. In case of the CsI analysis, the observed background count is taken to follow a Poisson distribution with a hypothetical parameter $N_{BG}$, that is marginalized. In the absence of clear uncertainties on the observed background count, this approach takes into account the statistical errors associated with this observation. Note that the uncertainties in each bin are marginalized over separately while building the likelihood function. This ensures that the signal shape information over all the free parameters ($\vec{\theta}$) is accounted for while comparing the predicted and observed data.

The likelihood function for the CsI data given model parameters ($\va*{\theta}$) is
\begin{eqnarray}
\mathcal{L}_{\rm CsI} \propto \prod_{\rm bins} && \int \dd \alpha \, \dd N_{BG} \, \mathcal{P}\left[N^{tot}_{obs},\lambda(\alpha,\va*{\theta})\right]\nn
&\times& \mathcal{N}\left[\alpha , 0, \sigma_\alpha\right] \times \mathcal{P}\left[N^{BG}_{obs},N_{BG}\right]\label{eqn:lklcsi}
\end{eqnarray}
where,
\begin{eqnarray}
\lambda(\alpha,\va*{\theta}) &=& (1+\alpha) N(\va*{\theta})+N_{bg}\nn
\mathcal{P}\left[k,\lambda\right]&=&\exp \{- \lambda\}\frac{\{\lambda\}^{k}}{k!}\nn
\mathcal{N}\left[x,\mu,\sigma\right] &=& \frac{1}{\sqrt{2\pi}}\frac{\exp (-x^2/2\sigma^2)}{\sqrt{\sigma^2}}
\end{eqnarray}
are the parameter for the Poisson distribution, the Poisson distribution probability mass function (PMF) and the Gaussian distribution PMF respectively. $N(\va*{\theta})$ is the \cens model prediction in each bin. The dummy variable $\alpha$ is used to marginalize the likelihood in each bin over the prediction uncertainty using a Gaussian envelope with standard deviation $\sigma_\alpha = 0.28$.

\begin{table}[b]
    \centering
    \begin{tabular}{|c|c|c|}
    \hline\hline
    &&\\
    & CENNS10 & Using $\mathcal{L}_{Ar}$\\ &&\\ \hline
    &&\\
    $N_{CE\nu NS}$& $159\pm43$&$159\pm41$\\ &&\\ \hline
    &&\\
    $N_{SS}$&$3131\pm23$&$3131\pm22$\\ &&\\ \hline
    &&\\
    $N_{pBRN}$& $553\pm34$&$555\pm33$\\ &&\\ \hline
    &&\\
    $N_{dBRN}$&$10\pm11$&$11\pm11$\\ &&\\ \hline \hline
    \end{tabular}
    \caption{Best-fit values and uncertainties on the \cens and associated background counts at the CENNS10 detector~\cite{Akimov:2020pdx}. The first column lists the values reported by the COHERENT Collaboration while the second column lists the values obtained using the likelihood function defined in Eq. \ref{eqn:lklar}}
    \label{tab:Arfits}
\end{table}

The likelihood function for the CENNS10 data is defined with the same basic principles in mind. Unlike the CsI data, it has significantly large BRN observations that must be taken into consideration. The CENNS10 Datarelease~\cite{Akimov:2020pdx} also provides the uncertainties associated with each of the three background components, that can now be incorporated directly. The likelihood function for the CENNS10 data is defined to be,
\begin{align}
&\mathcal{L}_{\rm Ar} \propto \nn &\prod_{\rm bins} \int \mathcal{P}\left[N^{tot}_{obs},\lambda({\alpha_i},\va*{\theta})\right]\left(\prod_i
\mathcal{N}\left[\alpha_i , 0, \sigma_{\alpha_i}\right]\, \dd \alpha_i\right).\label{eqn:lklar}
\end{align}
Here, the parameter $\lambda({\alpha_i},\va*{\theta})$ is
\begin{eqnarray}
    \lambda({\alpha_i},\va*{\theta}) &=& (1 + \alpha_1)\,N(\va*{\theta}) + (1 + \alpha_2)\,N_{SS}\nn &+& (1 + \alpha_3)\,N_{pBRN} + (1 + \alpha_4)\,N_{dBRN}),
\end{eqnarray}
where $N_{SS}$, $N_{pBRN}$, and $N_{dBRN}$ are the steady state, prompt BRN and delayed BRN background counts, respectively, in each bin. Just as in the case for CsI, the parameters, $\alpha_i$, are each the dummy variables that are marginalized to replicate uncertainties in each of the respective distributions as described in the first column of Table \ref{tab:Arfits}. Only the prediction uncertainty ($13\%$) is considered for the \cens count when estimating the model parameters.

In order to check the robustness of $\mathcal{L}_{Ar}$, we use it in repeating the likelihood maximization test described in Analysis A of Ref.\cite{Akimov:2020pdx}. Instead of marginalizing over the parameters $\alpha_i$, we pass them as parameters to be estimated by maximizing the likelihood function in Eq. \ref{eqn:lklar} without the Gaussian envelopes. We use the SM prediction of \cens for the values of $N(\va*{\theta})$ and a uniform prior on its normalization, $\alpha_1$. For the backgrounds, we use predicted counts instead of the best-fit counts as central values (as provided in Table I of Ref.\cite{Akimov:2020pdx}) and Gaussian priors on their normalizations with standard deviations
\begin{eqnarray}
    \sigma_{pBRN} &=& 0.3\nn
    \sigma_{dBRN} &=& 1.0.
\end{eqnarray}
The standard deviation for the prior on the steady state background normalization was taken to be the same as the uncertainty on its predicted value. The results of this table are shown in the second column of Table \ref{tab:Arfits}. For comparison, we also show the corresponding best-fit values and the errors as reported by the COHERENT Collaboration and we find them to be an excellent match.

Coming back to the estimation of model parameters using the likelihood functions defined in Eqs. \ref{eqn:lklcsi} and \ref{eqn:lklar}, the combined likelihood function was defined to be the product of the two separate ones,
\begin{equation}
    \mathcal{L}(\va*{\theta}) \propto \mathcal{L}_{CsI} \times \mathcal{L}_{Ar}.
\end{equation}
We use this likelihood function in two sets of maximization tests, one where $\va*{\theta}\equiv (g_x)$ with $M_{Z^\prime}$ as a fixed parameter and another where $\va*{\theta}\equiv (g_x,M_{Z^\prime})$. The ratio $r$ is taken as a fixed parameter for both analyses. We always use noninformative priors on the estimated parameters $(\va*{\theta})$ and maximize the likelihood using the MultiNest algorithm~\cite{Feroz:2008xx}.

We use the log-ratio test~\cite{wilks1938} to calculate the deviation of the maximum likelihood estimate (MLE) from the SM or the experimental best-fit observation while evaluating exclusions. In comparisons of two different models with different sets of estimated parameters, we define the test statistic,
\begin{equation}
    t = -2\, \log\left(\frac{\mathcal{L}_0 }{\mathcal{L}(\vu*{\theta})}\right).
\end{equation}
\begin{figure*}[t]
    \centering
    \subfigure[\label{fig:gxm20}]{\includegraphics[width=8cm]{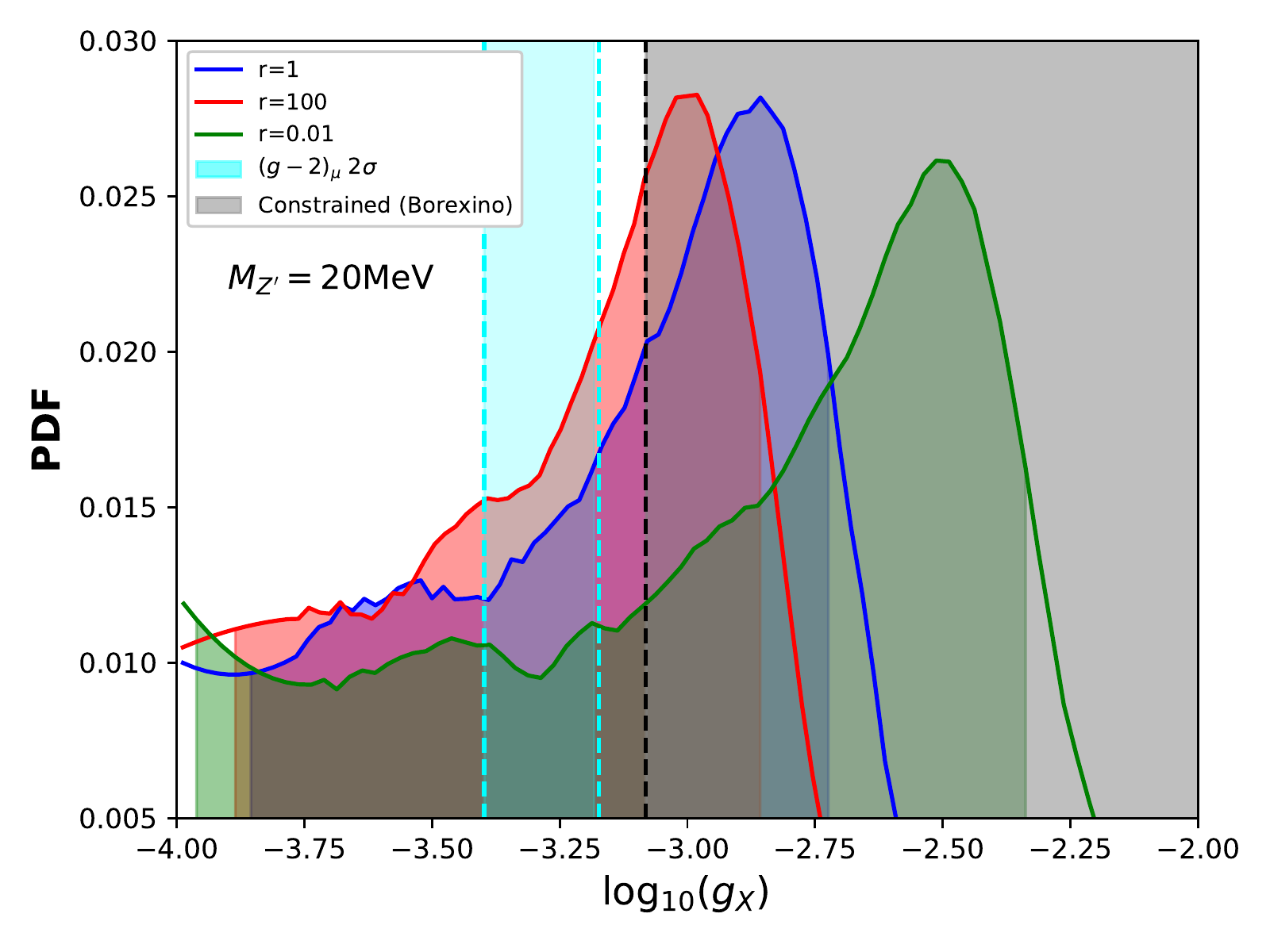}}
    \subfigure[\label{fig:gxm100}]{\includegraphics[width=8cm]{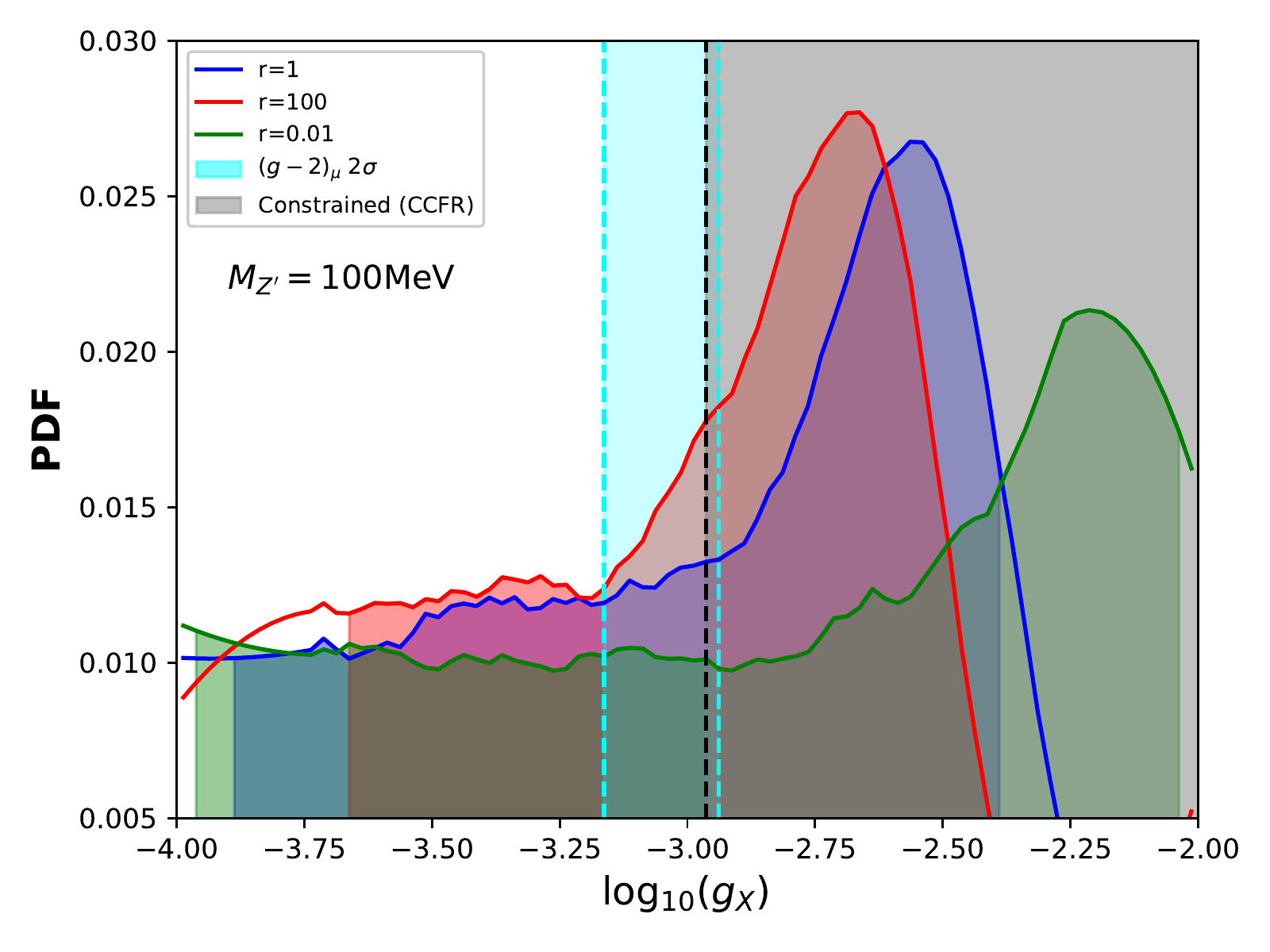}}
    \caption{The probability distribution functions estimating $g_x$ for fixed values of $M_{Z^\prime}$ and $r$ from the likelihood maximization test involving CsI(2017) + CENNS10 data from the COHERENT Collaboration. For perspective, we also show the range of $g_x$ that can explain $(g-2)_\mu$ (cyan), as well as those values for it that are constrained by experiments like CCFR~\cite{Mishra:1991bv}, Borexino~\cite{Bellini:2013lnn} and BaBar~\cite{TheBABAR:2016rlg} (gray). The $1\sigma$ bands for each of the $g_x$ PDFs are shaded in the color of the respective curves. The SM is excluded by the $1\sigma$ bands from the likelihood profile. Larger values of $M_{Z^\prime}$ prefer larger values of $g_x$. Larger values of $r$ prefer smaller values of $g_x$, have a marginally higher likelihood and have a smaller estimation uncertainty.}
    \label{fig:gxmlkl}
\end{figure*}
For our purposes, $\mathcal{L}_0$ represents the SM likelihood with no estimated parameters and $\mathcal{L}(\vu*{\theta})$ is the likelihood at the MLE. According to Wilks' theorem this test statistic is distributed as a $\chi^2_\eta$ distribution where $\eta$ is the difference between the number of estimated parameters in the two models being compared. Standard procedure can then be followed to determine the $p$-value ($p= {\rm Pr}\left[\chi^2_\eta \, > \, t\right]$) for this test and the corresponding significance,
\begin{equation}
    Z = \Phi^{-1}(1-p/2)
\end{equation}
where $\Phi^{-1}$ is the inverse cumulative distribution function of the standard normal distribution. Significant results may be tested by either a small p-value or large values of Z.
For calculating exclusions we use
\begin{equation}
    t = -2\, \log\left(\frac{\mathcal{L}(\va*{\theta}) }{\mathcal{L}_1}\right)
\end{equation}
where $\mathcal{L}_1$ is the likelihood of the best-fit experimental observation. Values of parameters entering $\va*{\theta}$ that result in $Z>2$ using this test statistic are then excluded at a $2\sigma$ C.L.

\section{Results}
\begin{figure*}[t]
    \centering
    \subfigure[\label{fig:gxm20-chi}]{\includegraphics[width=8cm]{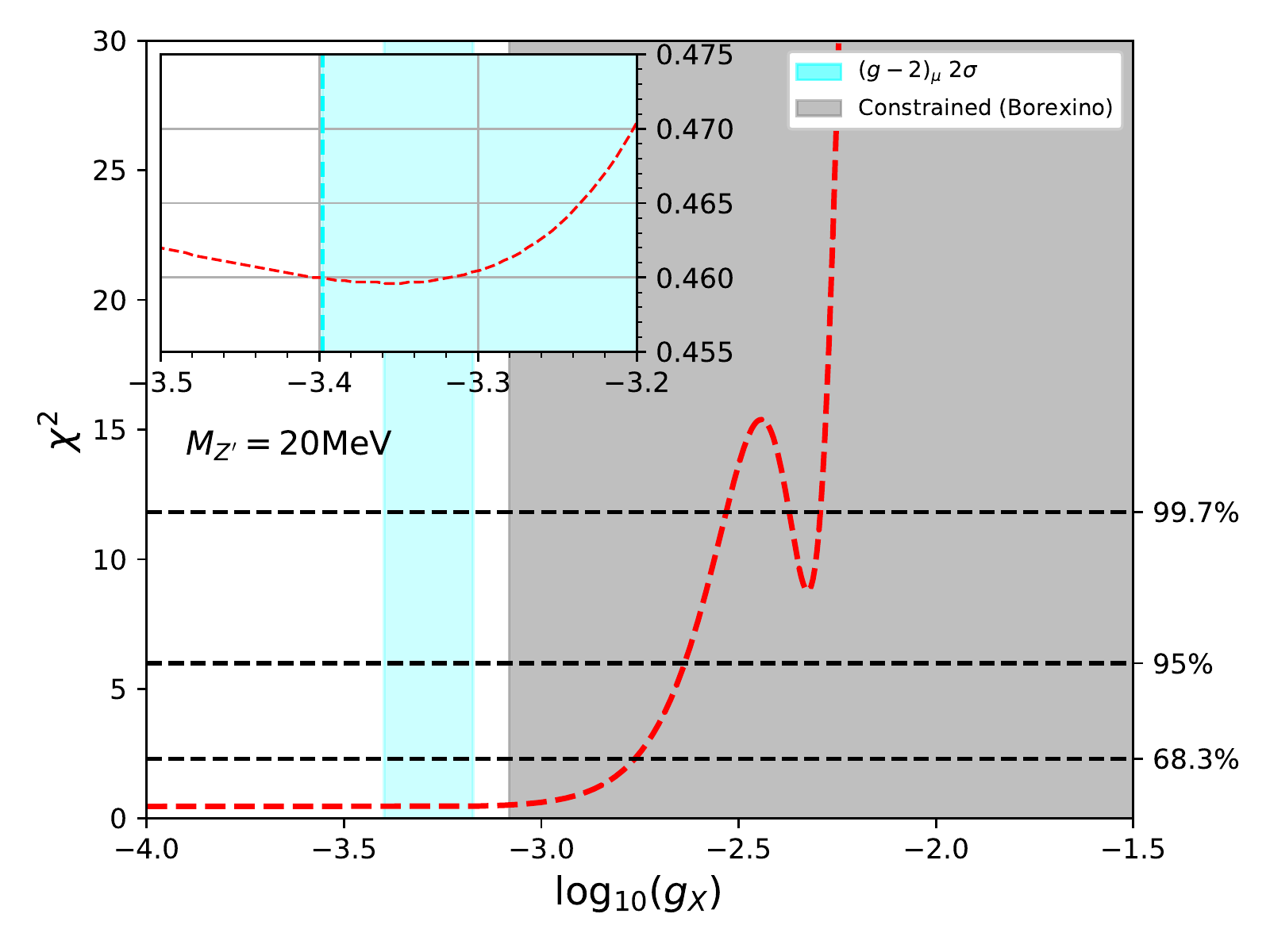}}
    \subfigure[\label{fig:gxm100-chi}]{\includegraphics[width=8cm]{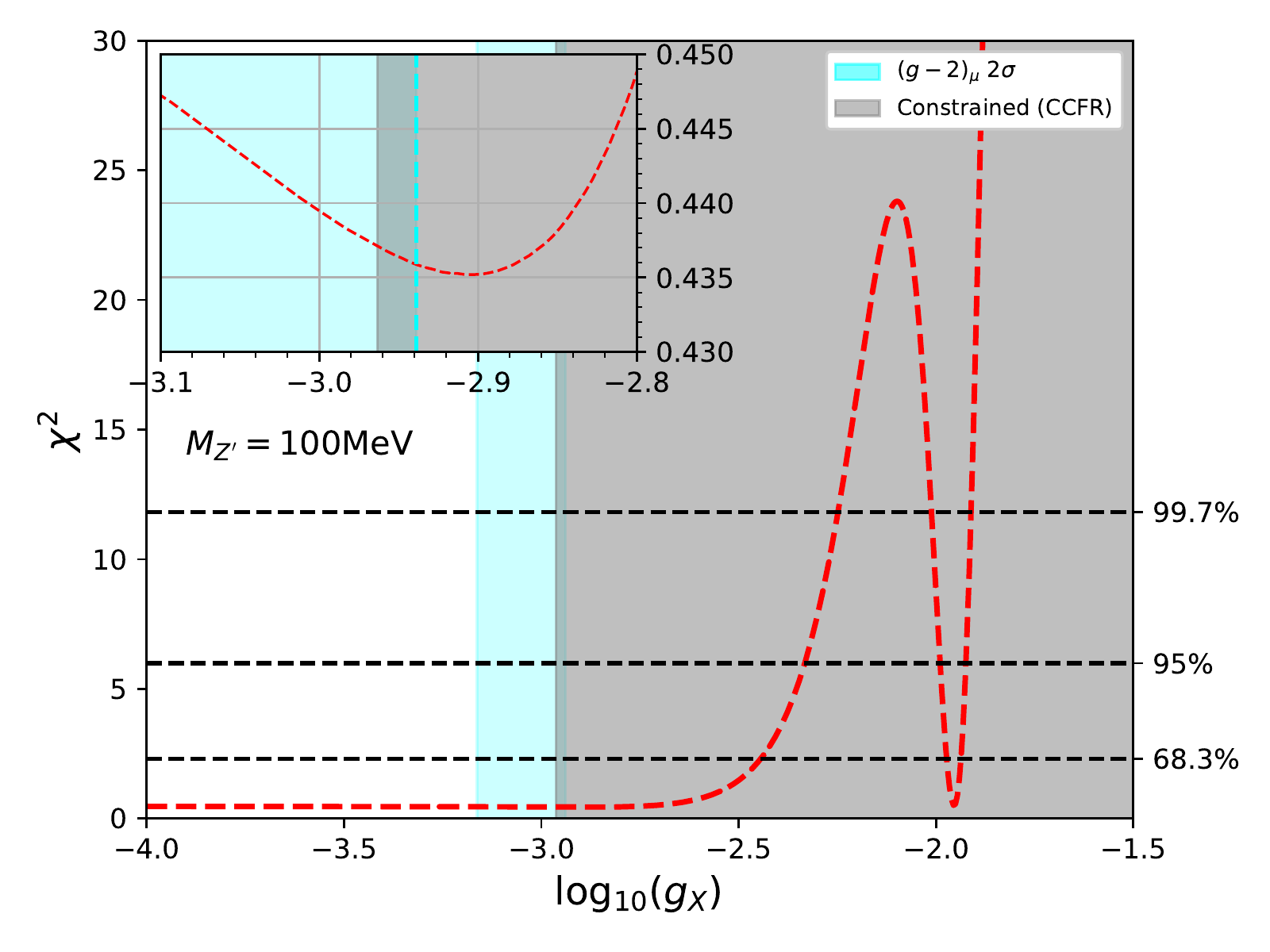}}
    \caption{Variation of the chi-square deviation as defined in Eqs. \ref{eqn:chi-csi} and \ref{eqn:chi-ar} with $g_x$ for fixed values of $M_{Z^\prime}$ and $r=1$ using CsI(2020)+CENNS10 data from the COHERENT Collaboration. The insets show the global minima of the chi-square profile which are for nonzero values of the additional gauge boson couplings. Larger choices of $M_{Z^\prime}$ beget larger values of preferred $g_x$ and also lead to an additional region of preference at large $g_x$. For perspective, we also show the range of $g_x$ that can explain $(g-2)_\mu$ (cyan), as well as those values for it that are constrained by experiments like CCFR~\cite{Mishra:1991bv}, Borexino~\cite{Bellini:2013lnn} and BaBar~\cite{TheBABAR:2016rlg} (gray). Only the unbinned total event counts from the detectors are used for this test. The SM is within $1\sigma$ of the best fit value of $g_x$.}
    \label{fig:gxm}
\end{figure*}

\begin{figure*}[t]
    \centering
    \subfigure[\label{fig:combr001}]{\includegraphics[width=8cm]{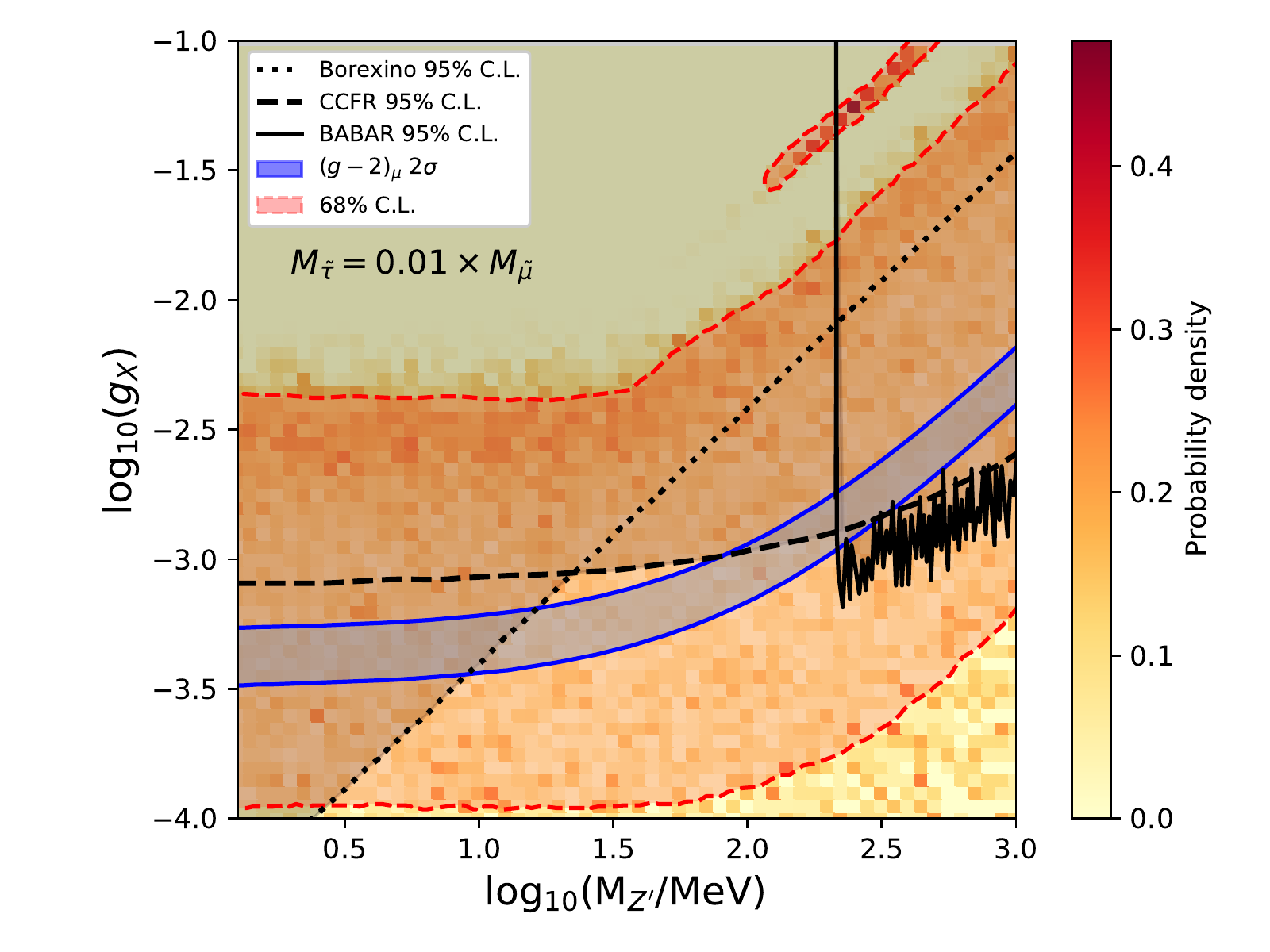}}
    \subfigure[\label{fig:combr1}]{\includegraphics[width=8cm]{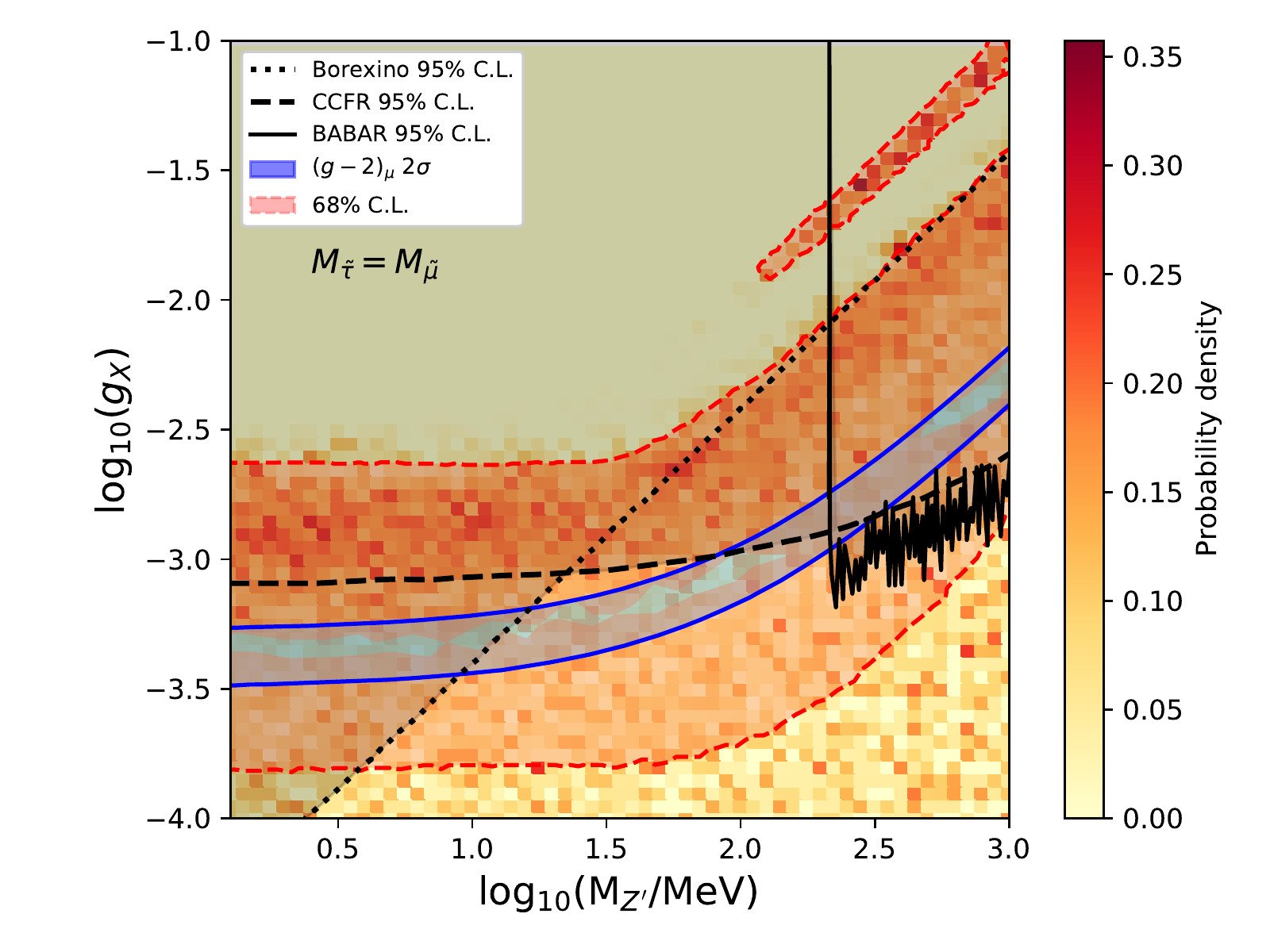}}
    \subfigure[\label{fig:combr100}]{\includegraphics[width=8cm]{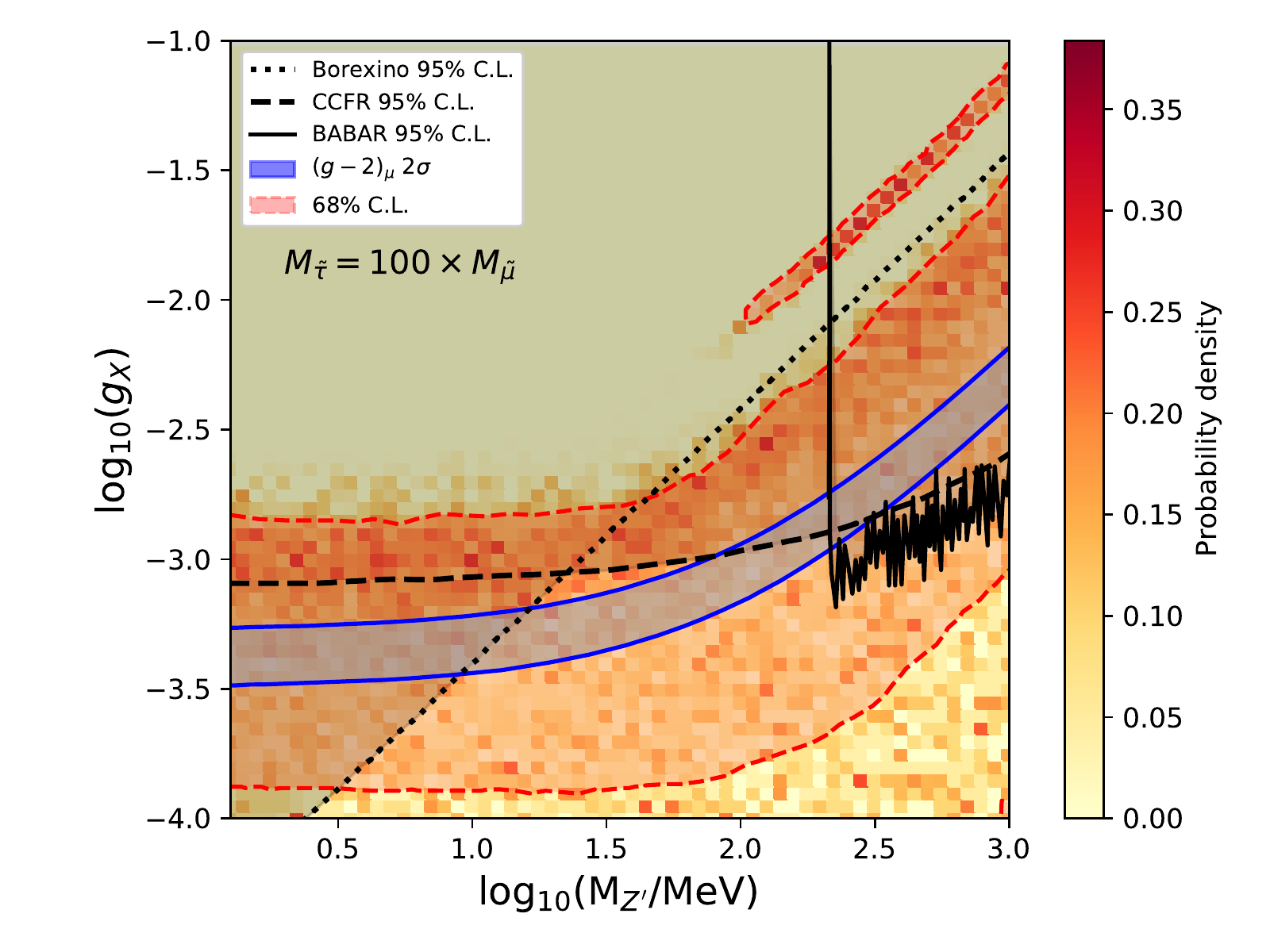}}
    \caption{Heat maps showing probability distribution from the maximum likelihood estimation in the $g_x-M_{Z^\prime}$ plane using CsI(2017) + CENNS10 data from the COHERENT Collaboration. Parameter space preferred within a $68\%$ C.L. are shown shaded in red within the red dashed lines. We also show the parameter space that can explain $(g-2)_\mu$ (blue), as well as the region constrained by experiments like CCFR~\cite{Mishra:1991bv}, Borexino~\cite{Bellini:2013lnn} and BaBar~\cite{TheBABAR:2016rlg} (gray).}
    \label{fig:heatmap}
\end{figure*}
\begin{figure}[t]
    \centering
  \includegraphics[width=8cm]{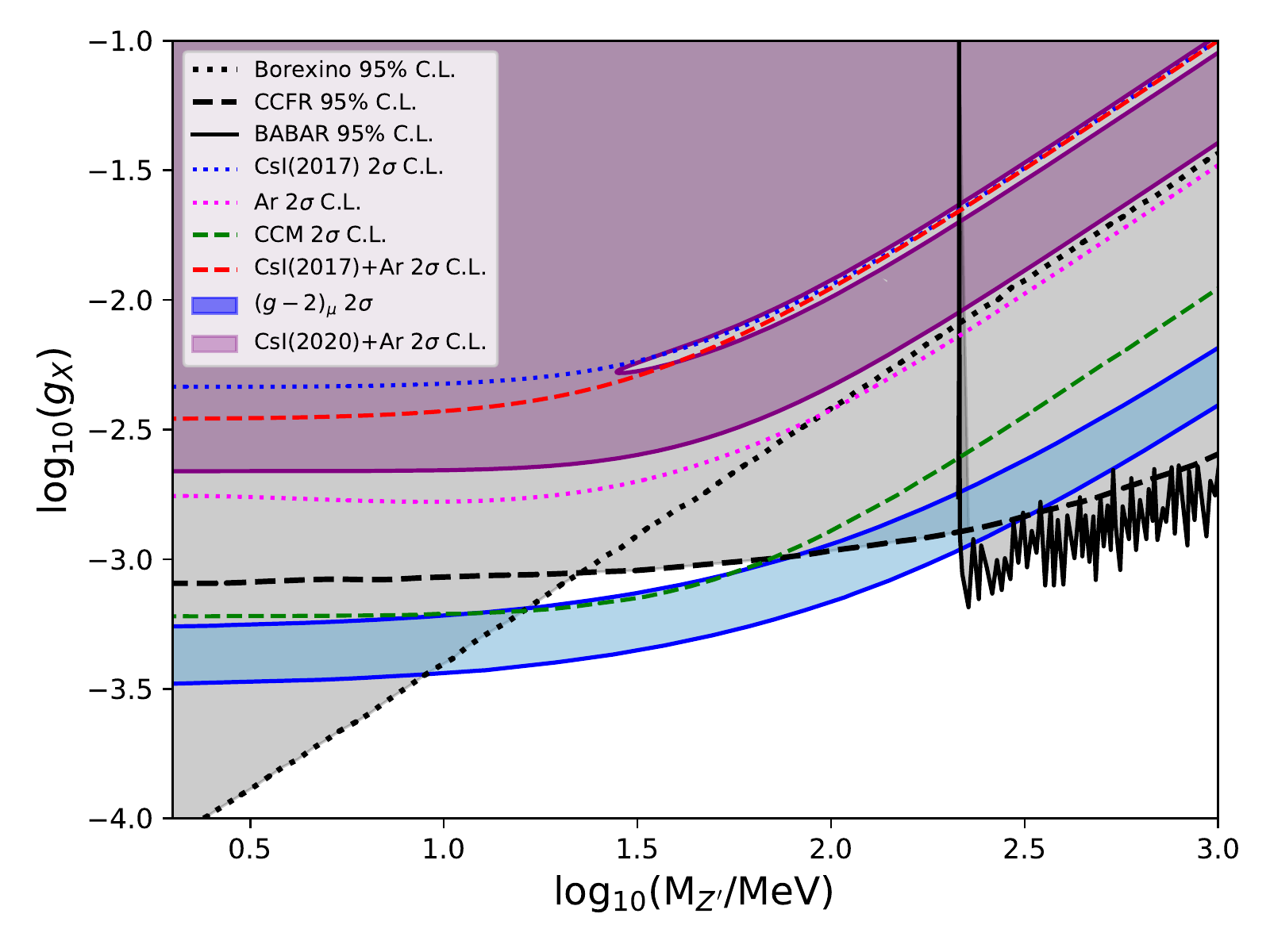}
    \caption{Exclusions from the likelihood based analysis of CsI(2017), CENNS10 and their combination on the $L_\mu-L_\tau$ parameter space are shown. Projected exclusions from the CCM experiment assuming three years of data taking are also shown. The exclusions shown are for $r=1$, which also corresponds to the non-SUSY case. Larger values of $r$ would lead to stronger constraints. The exclusions on the same parameter space from the analysis of the chi-square deviation using the CENNS10 and updated CsI(2020) total counts are also shown.}
    \label{fig:exclusion}
\end{figure}
\begin{table}[b]
    \centering
    \begin{tabular}{|c|c|c|c|}
    \hline \hline 
    &&&\\
    $M_{Z^\prime}$& r=0.01 & r=1.0 & r=100\\
    &&&\\
    \hline
    &&&\\
    20 MeV&$(0.1-4.87)$&$(0.13-1.99)$&$(0.12-1.46)$\\
    & $1.40\sigma$ & $1.40\sigma$ & $1.40\sigma$\\
    &&&\\
    \hline
    &&&\\
    100 MeV&$(0.1-9.7)$&$(0.12-4.3)$&$(0.2-7.7)$\\
    & $1.32\sigma$ & $1.33\sigma$ & $1.33\sigma$\\
    &&&\\
    \hline \hline
    \end{tabular}
    \caption{The $1\sigma$ limits of the estimates for $g_x$ in units of $10^{-3}$ along with the corresponding deviations of the MLE from the SM. These are the results of the likelihood maximization test for all the chosen values of $M_{Z^\prime}$ and $r$ using combined CsI(2017) and CENNS10 data from the COHERENT Collaboration.}
    \label{tab:combfit}
\end{table}

We begin with the likelihood maximization test where $g_x$ is estimated by fixing the values of both $M_{Z^\prime}$ and $r$. The results are shown in Fig. \ref{fig:gxmlkl}. The $1\sigma$ C.L. regions are shaded in the color of the corresponding PDF curves. We repeat the test for two separate values of $M_{Z^\prime}$, 20 MeV [Fig.\ref{fig:gxm20}] and 100 MeV [Fig.\ref{fig:gxm100}] that represent two choices on either ends of the unconstrained range of values that are preferred by $(g-2)_\mu$. They also show the range of $g_x$ favored by the $(g-2)_\mu$ anomaly for each value of $M_{Z^\prime}$ and the region disfavored by experiments like CCFR~\cite{Mishra:1991bv}, Borexino~\cite{Bellini:2013lnn}, and BaBar~\cite{TheBABAR:2016rlg}. The $r=1$ lines correspond to the non-SUSY model (which in this case is the same as the case for degenerate sleptons). The effect of nonvanishing contribution from SUSY sources is captured entirely by choosing other values of $r$. We show the results for two representative values of $r$ besides the $r=1$ result, $r=0.01$ and $r=100$. These choices capture a wide range of the most common choices for the slepton mass hierarchies.

For all choices of the fixed parameters, we find that the range of $g_x$ favored by $(g-2)_\mu$ is entirely within $1\sigma$ of the MLE while the SM is beyond the $1\sigma$ C.L. Also, larger values of the stau to smuon mass ratio ($r$) are slightly preferred over their degeneracy (which is the same as the absence of SUSY in this case) or smaller values. They also have narrower uncertainties, which denotes a better fit to data, and a better overlap with the $(g-2)_\mu$ range. All the $1\sigma$ intervals of the fit estimates for $g_x$ and the deviation of the corresponding MLE from the SM are shown in Table~\ref{tab:combfit}.

The $g_x$ dependence of the chi-square deviation as described in Sec. \ref{sec:csi-update} for fixed values of $M_{Z^\prime}$ and $r=1$ are shown in Fig. \ref{fig:gxm}. The chi-square deviation is defined using total counts only as the full energy-time binned CsI(2020) data is still not in the public domain. The insets in the figures show the variation of the function around its global minima. As in the previous test, we find that the global minima coincides with the $(g-2)_{\mu}$ preferred parameter range for the choices of 20 MeV$\lesssim M_{Z^\prime}\lesssim$100 MeV. The global minima of the chi-square profile is at $3.2\times 10^{-4}$ and $1.3\times 10^{-3}$ for $M_{Z^\prime}=$ 20MeV and 100MeV, respectively, which are within the range of couplings allowed from the $(g-2)_{\mu}$ observations. This implies that the band of gauge boson masses from $20-100$ MeV better fits the observations of the COHERENT Collaboration than the SM while also explaining $(g-2)_{\mu}$ in tandem. There is an additional minima for larger values of $g_x$ that becomes the global minima for  $M_{Z^\prime}\gtrsim 200$ MeV but these choices are already ruled out by observations from the experiments mentioned earlier. In contrast to the likelihood profiles, the $\chi^2$ test does not exclude the SM at the $1\sigma$ C.L. This distinction is most important as the likelihood test uses both energy and timing distributions from both detectors while the $\chi^2$ test uses only the total counts. However the latter also includes the latest CsI(2020) data while the former does not. Whether the full updated data in a likelihood analysis still has the preference for BSM physics will be clear once the Collaboration makes the data public.

We also show the results of the likelihood maximization test where both $g_x$ and $M_{Z^\prime}$ are estimated parameters and the ratio $r$ is fixed. The results of this analysis are summarized in Fig.\ref{fig:heatmap} where the three different panels demonstrate the effect of three different choices of $r$. Once again, we use three representative values, $r=0.01$ [Fig. \ref{fig:combr001}], $r=1$ [Fig. \ref{fig:combr1}], and $r=100$ [Fig.\ref{fig:combr100}]. The color code denotes the variation of the PDF. Since the MultiNest algorithm rejects parameter points over iterations by importance sampling, regions of lower likelihood also have a lower density of final reported likelihood points. The red shaded region in these figures denote the parameter space preferred by the COHERENT data within $1\sigma$. We also show the $(g-2)_{\mu}$ preferred and already constrained parameter space in these figures.

There are a few features of these results that are common across all the choices of $r$.  The SM ($g_x = 0$) is always within $2\sigma$ of the MLE, however there is a slight preference for nonzero $L_\mu-L_\tau$ gauge coupling. Although the MLE is always in regions of the parameter space that have already been excluded by experiments like CCFR~\cite{Mishra:1991bv}, Borexino~\cite{Bellini:2013lnn} and BaBar~\cite{TheBABAR:2016rlg}, unconstrained regions of the parameter space where the $(g-2)_{\mu}$ may be explained are within $1\sigma$ of it. Changing the value of $r$ affects the estimated most likely value of $g_x$. Larger values of $r$ prefer smaller values of $g_X$ and vice versa. Larger values of $r$ also bring the preferred values of $g_x$ into better overlap with those favored by $(g-2)_\mu$. At the same time, larger values of this ratio are also associated with a slightly larger likelihood. However, the logarithmic dependence of $\epsilon$ on $r$ ensures that these dependencies are quite mild as we go for even larger values than 100.

The exclusions from these analyses using the method outlined in Sec. \ref{sec:lkltest} are shown in Fig. \ref{fig:exclusion}. The exclusions from the CENNS10 analysis are far stronger than those from CsI(2017). This is because it observed an excess of events above the SM while the ($L_\mu-L_\tau$) $Z^\prime$ serves to reduce the count for the relevant parameter space. The CsI(2017) data alone has a $\sim2\sigma$ preference for a $Z^\prime$ coupling exclusively to second generation leptons~\cite{Dutta:2019eml}, leading to a smaller \cens count than the SM prediction. Although this preference has a marked reduction in the combined data, it does survive. As a result, the exclusions from the combined data closely follow those from CsI(2017) alone. These exclusions have been shown keeping the ratio $r=1$ fixed. Since larger ratios show a preference for smaller values of $g_x$, they also correspond to stronger exclusions. However, as mentioned earlier, the dependence of the kinetic mixing on this ratio is logarithmic. Hence, the change in the exclusions get smaller and smaller for larger values of $r$. The $95\%$ confidence level exclusion from the chi-square profile on the $g_x$-$M_{Z^\prime}$ plane is also shown in Fig. \ref{fig:exclusion}. These constraints are stronger than the existing ones from the \cens detection by the Collaboration, however they are not the strongest in the relevant parameter space. 

In addition we also show the projected exclusion from the CCM experiment. Since we have no information for the backgrounds for this experiment yet, we assume that the \cens count itself has a Poisson distribution in each bin. We also use a detector threshold of 25 keV and above that we use $100\%$ efficiency and a quenching factor of 1PE/keV for easy scalability once the information is available. We have used signal predictions using a projected live-period of $\sim5000$hrs/year for three years which correspond to $\sim3\times10^{22}$ protons on target. We modify the likelihood function in Eq. \ref{eqn:lklar} suitably by setting the background counts to zero. We also have no information on the associated systematic uncertainties for it, and hence we do not use the dummy variables for marginalization over errors. The only possible error that could be included is the $\sim2\%$ theoretical error on the \cens prediction. However these are too small to cause any significant change to our results and hence are disregarded. Assuming that the CCM experiment observes a \cens count equal to the SM prediction, it could potentially exclude hitherto unconstrained portions of the $L_\mu-L_\tau$ parameter space. More importantly, it may be able to exclude a portion of the parameter space favored by the $(g-2)_\mu$ observations. Similar sensitivities in constraining neutrino NSI from the CCM experiment have also been observed by the authors of Ref.~\cite{Shoemaker:2021hvm}.

\section{Conclusion}
We undertook a comprehensive statistical analysis of the combined CsI and CENNS10 data recently made public by the COHERENT Collaboration. The full energy and timing data from the CsI detector as well as the energy, timing and PSD data from the CENNS10 detector was used to estimate the best-fit parameters for the $L_\mu-L_\tau$ model and their deviation from the SM prediction. We have also studied the effect that the presence of SUSY may have on these results.

We find a slight preference for reductive BSM effect from the chi-square deviation analysis involving total counts of the updated CsI and CENNS10 data from the COHERENT Collaboration. The global minima of the chi-square profile for $20{\rm MeV} \lesssim M_{Z^\prime} \lesssim 100{\rm MeV}$ are for values of coupling $g_x$ that can explain $(g-2)_{\mu}$ for the respective $Z^\prime$ masses and are also unconstrained. An additional minima for larger values of $g_x$ is distinct especially for a heavier $Z^\prime$. This additional minima becomes the global minima for $M_{Z^\prime}\gtrsim 200$ MeV, however the choice of this parameter space is constrained by the observations of the CCFR and BaBar Collaborations. A $\sim 1.3\sigma$ preference for reductive BSM effect as well as an overlap with the $(g-2)_{\mu}$ favored parameter space within $1\sigma$ was also observed from the likelihood profile of the CsI(2017)+ CENNS10 energy, timing and PSD distribution. An important distinction between the two different analyses is that the likelihood profile has no overlap with the SM at the $1\sigma$ C.L. In contrast, the $\chi^2$ test using only the total counts from CsI(2020)+CENNS10 data has an overlap with the SM within $1\sigma$ despite the global minima mildly preferring nonzero new physics contribution.

The exclusions from the CENNS10 data that observed an excess of counts over the SM \cens prediction are much stronger than the earlier CsI exclusions on the $L_\mu-L_\tau$ parameter space. However, the surviving preference for reductive BSM effects results in the exclusions from the combined data to closely adhere to the CsI ones. The exclusions from the combined CENNS10 and CsI(2020) total counts using a $\chi^2$-test are stronger than the CsI(2017) + CENNS10 exclusions, but are still weaker than those coming from the CENNS10 data alone. Still, neither the CENNS10 exclusions by themselves or those from the combined data are the strongest exclusions as far as the $L_\mu-L_\tau$ model is concerned. We also show the projected exclusions from the CCM experiment. Assuming it observes the SM prediction for \cens, it might be able to exclude yet unconstrained regions of the parameter space favored by the $(g-2)_\mu$. 

Some of the other upcoming \cens experiments also look promising in this regard. For instance, the planned 610 kg extension (CENNS610) of the CENNS10 LAr detector at the SNS would be able to significantly improve statistics on the existing data. The experiments at ESS and especially ESS10~\cite{Baxter:2019mcx} with its strikingly low threshold energy are expected to probe large parts of the model parameter space including those that can explain the present $(g-2)_{\mu}$ measurements. Other experiments like the NA64$\mu$\cite{Gninenko:2653581,Gninenko:2640930} muon beam experiment, the NA62\cite{Krnjaic:2019rsv} Kaon factory  and liquid Xenon direct Dark Matter detectors like the LZ\cite{Mount:2017qzi}, XENONnT\cite{Aprile:2020vtw} and the DARWIN observatory\cite{Aalbers:2016jon}  also have promising capabilities to probe regions of the \newu parameter space pertinent to the $(g-2)_\mu$. For a recent study of projected constraints from these experiments on the $L_\mu-L_\tau$ parameter space see, for example, Ref.\cite{Amaral:2021rzw}. Observation of neutrino tridents at the DUNE near detector is also projected to be able to improve upon the CCFR observations and probe the $(g-2)_\mu$ favored parameter space of the $L_\mu-L_\tau$ model\cite{Altmannshofer:2019zhy,Abi:2020kei}.

The formalism for the SUSY scenario was kept the same as that in Ref.~\cite{Banerjee:2020zvi}. Our analysis finds that the range of $g_x$ favoured by $(g-2)_{\mu,e}$ in this scenario overlaps with the estimates of $g_x$ from the likelihood maximization within $1\sigma$. We also find that larger values of the ratio $r$ have better overlap with this range of $g_x$ and is slightly more favored by the combined data.

This exercise also points towards the fact that the combined COHERENT data still prefers BSM effects that serve to reduce the \cens count predicted from the SM, although to a lesser degree than with the CsI data alone. The CsI(2017) data preferred reductive BSM effects via light mediators at $\sim2\sigma$~\cite{Dutta:2019eml}. The preference for similar effects in the combined data from CsI(2017) + CENNS10 has been reduced to $\sim 1.3-1.4\sigma$.

\begin{acknowledgments}
H.B. thanks Shu Liao for helpful discussions about the analysis of the CsI data and Samiran Sinha for his valuable insights on the likelihood function. We thank A. Thompson, L. Strigari, G. Rich and D. Pershey for very useful discussions regarding the analysis of the CENNS10 data. We also thank D. Pershey for a careful reading of the manuscript. S.R. acknowledges the hospitality of the Mitchell Institute for Fundamental Physics and Astronomy, Texas A \& M University, during the initial period of this work. The work of B.D. is supported in part by the DOE Grant No. DE-SC0010813. 
\end{acknowledgments}
%


\end{document}